\documentclass[journal]{IEEEtran}

\usepackage{multirow}

\usepackage{amsmath}
\usepackage{amsfonts,amsthm,amssymb}
\DeclareMathOperator{\sinc}{sinc}

\usepackage{verbatim}

\usepackage[caption=false,font=footnotesize]{subfig}
\usepackage{mathtools}
\usepackage{nicefrac}
\usepackage{bm}

\usepackage{balance}

\usepackage{upquote}
\usepackage[none]{hyphenat}
\usepackage{graphicx}

\usepackage{cite}

\usepackage{url}

\usepackage[inline]{enumitem}
\usepackage[normalem]{ulem}
\usepackage{xcolor}
\newcommand\redsout{\bgroup\markoverwith{\textcolor{red}{\rule[0.5ex]{2pt}{0.4pt}}}\ULon}

\def\IEEEbibitemsep{0pt plus .4pt}
\begin{document}

\title{Index-Modulated Metasurface Transceiver Design using Reconfigurable Intelligent Surfaces for 6G Wireless Networks}

\author{John~A.~Hodge$^\ast$,~\IEEEmembership{Student~Member,~IEEE,}
        Kumar~Vijay~Mishra$^\ast$,~\IEEEmembership{Senior~Member,~IEEE,} Brian M. Sadler~\IEEEmembership{Life Fellow,~IEEE}, Amir~I.~Zaghloul,~\IEEEmembership{Life~Fellow,~IEEE}
\thanks{$^\ast$J. A. H. and K. V. M. are co-first authors.}        
\thanks{J. A. H. is with Bradley Department of Electrical and Computer Engineering, Virginia Tech, Blacksburg, VA 24061 USA. Email: jah70@vt.edu.}
\thanks{K. V. M. and B. M. S. are with the United States DEVCOM Army Research Laboratory, Adelphi, MD 20783 USA. E-mail: kvm@ieee.org, brian.m.sadler6.civ@army.mil.}
\thanks{A. I. Z. is with Bradley Department of Electrical and Computer Engineering, Virginia Tech, Blacksburg, VA 24061 USA, and United States DEVCOM Army Research Laboratory, Adelphi, MD 20783 USA. E-mail: amirz@vt.edu.}
\thanks{J. A. H. acknowledges support from Northrop Grumman Mission Systems (NGMS), Baltimore, MD, for his thesis research. K. V. M. acknowledges support from the National Academies of Sciences, Engineering, and Medicine via the Army Research Laboratory Harry Diamond Distinguished Postdoctoral Fellowship. Research was sponsored by the Army Research Laboratory and was accomplished under Cooperative Agreement Number W911NF-21-2-0288. The views and conclusions contained in this document are those of the authors and should not be interpreted as representing the official policies, either expressed or implied, of the Army Research Laboratory or the U.S. Government. The U.S. Government is authorized to reproduce and distribute reprints for Government purposes notwithstanding any copyright notation herein.}
\thanks{The conference precursor of this work was presented at the 2021 IEEE International Conference on Acoustics, Speech, and Signal Processing (ICASSP).}
}

\maketitle

\begin{abstract}
Higher spectral and energy efficiencies are the envisioned defining characteristics of high data-rate sixth-generation (6G) wireless networks. One of the enabling technologies to meet these requirements is index modulation (IM), which transmits information through permutations of indices of spatial, frequency, or temporal media. In this paper, we propose novel electromagnetics-compliant designs of reconfigurable intelligent surface (RIS) apertures for realizing IM in 6G transceivers. We consider RIS modeling and implementation of spatial and subcarrier IMs, including beam steering, spatial multiplexing, and phase modulation capabilities. Numerical experiments for our proposed implementations show that the bit error rates obtained via RIS-aided IM outperform traditional implementations. We further establish the programmability of these transceivers to vary the reflection phase and generate frequency harmonics for IM through full-wave electromagnetic analyses of a specific reflect-array metasurface implementation. 
\end{abstract}

\begin{IEEEkeywords}
6G wireless communications, coded metasurface, index modulation, reconfigurable intelligent surfaces, space-time modulated metasurface.
\end{IEEEkeywords}

\IEEEpeerreviewmaketitle

\section{Introduction}
\label{sec:intro}
\IEEEPARstart{A}{s} global demand for connectivity increases with a surge in mobile and Internet-of-things (IoT) devices, high power consumption and quality-of-service (QoS) in congested dense urban environments have become challenges for wireless network operators \cite{sisinni2018industrial}. To meet these demands of high data rates in fifth-generation (5G) and upcoming sixth-generation (6G) networks while also reducing power consumption, there is a need for wireless communications devices that are both energy and spectrally efficient (EE/SE) \cite{qi2019outage}. For better connectivity, future wireless networks need to mesh disparate systems \cite{sisinni2018industrial}, provide ultra-reliability to autonomous systems \cite{dokhanchi2019mmwave}, and intelligently adapt to changing environments by overcoming signal attenuation from multi-path fading, line-of-site blockages, and other uncontrollable interference \cite{elbir2022rise}.

It is, therefore, imperative to employ reconfigurable antennas that have the ability to emit more than one radiation pattern at different frequencies and polarizations, efficiently use the constrained spectrum, beamform for multiple-input multiple-output (MIMO) systems, and adaptively shape the beam at reduced cost and form factor \cite{christodoulou2012reconfigurable,hong2017multibeam}. In this context, reconfigurable/large intelligent surfaces (RIS/LIS) \cite{huang2019reconfigurable,hodge2019reconfigurable,mishra2019reconfigurable} have captured significant research interest in the wireless communications community. The RISs are capable of applying customized transformations to radio waves based on the propagation channel and provide low-cost multi-function integrated antennas and transceivers \cite{basar2020reconfigurable}.

In particular, systems that employ large arrays such as massive MIMO communications \cite{mishra2019toward} suffer from prohibitive costs and area associated with many radio-frequency (RF) chains. The RF module of each radiating element or sub-array typically comprises a power amplifier, mixer, phase shifters, and analog filters. The significant cost, complexity, and associated size-weight-and-power-cooling (SWaP-C) of these RF components preclude low-cost and energy-efficient solutions for many wireless applications. In this context, a RIS directly performs signal modulation on the antenna surface while combining some of the RF circuity with the programmable aperture \cite{basar2019wireless}. 

Another significant hindrance to future wireless systems arises from signaling strategies. Current modulation techniques like MIMO orthogonal frequency-division multiplexing (MIMO-OFDM) are not sufficient to satisfy 5/6G data requirements \cite{basar2020reconfigurable}. Recent studies suggest employing index modulation (IM) \cite{ishikawa201850} to achieve high SE and EE \cite{cheng2018index}, superior bit error rates (BERs) \cite{basar2020reconfigurable}, and robustness against inter-channel interference (ICI) and static fading \cite{mao2018novel}. In IM, information is transmitted through permutations of indices of spatial, frequency, or temporal media. Common IM techniques include spatial modulation (SM) and frequency subcarrier index modulation (SIM); the latter includes OFDM. Very recently, LIS-based spatial IM has been proposed as a new paradigm for beyond-MIMO 6G networks \cite{basar2020reconfigurable}. In this paper, we focus on designing RIS for various IM-based communications systems.

RIS is made up of a \textit{metasurface} (MTS), which is a two-dimensional (2-D) reconfigurable electromagnetic (EM) layer composed of a large periodic array of subwavelength scattering elements (meta-atoms) with specially designed spatial features \cite{holloway2012overview}. Specifically, we consider implementing RIS/LIS via reconfigurable MTSs (R-MTSs) \cite{mishra2019reconfigurable,hodge2019reconfigurable}. The R-MTSs are capable of controlling and manipulating EM waves through modified surface boundary conditions. These surfaces are electrically thin and comprise an array of spatially varying sub-wavelength scattering elements (or meta-atoms). Through careful engineering of each meta-atom \cite{hodge2019multi,hodge2019rf,hodge2019joint}, MTSs can transform an incident EM wave into an arbitrarily tailored transmitted or reflected wavefront. Recent developments in time-modulated metasurfaces have unlocked a new class of nonlinear and nonreciprocal behaviors including direct modulation of carrier waves, programmable frequency conversion, and controllable frequency harmonic generation \cite{hodge2019reconfigurable,zhang2018space,wu2019serrodyne}.

Realistic and EM-compliant RIS implementations are imperative for the adoption of future RIS-based wireless communications. While many recent studies have proposed RISs \cite{liaskos2018new,huang2019reconfigurable} and LISs \cite{basar2019wireless,basar2020reconfigurable,han2019large} for 5G and 6G communications, specific implementations and EM analysis of RIS designs remain relatively unexamined in the current literature. In \cite{liu2019intelligent}, EM modeling of intelligent MTSs was performed but realistic feed structures and specific adaptation for wireless communications were not accounted for. In this paper, we design R-MTS for IM-based systems, wherein the RF aperture and transceiver are integrated within the MTS. Our dynamic reconfiguration of the MTS aperture in a wireless communications transmitter (Tx) facilitates beam steering, frequency agility, and phase modulation without conventional front-end devices such as phase-shifters, mixers, and switches. We synthesize a practical realization of a programmable RF metasurface operating as an RIS for IM and spatial multiplexing. Then, our full-wave EM experiments demonstrate R-MTS operation for both SM and SIM signaling.

Preliminary results of our work appeared in our conference publications \cite{hodge2021performance,hodge2020media}. In this work, our main contributions are:\\
\textbf{1) EM-compliant circuit design for IM-based RIS.}  This work ties together wireless communication theory with EM analysis to demonstrate the potential of low-cost R-MTSs for implementing IM techniques for 5G and beyond wireless networks.\\
\textbf{2) IM-based RIS across several domains.} We present RIS-based designs for IM across all four major wireless media resources: spatial, spectral, temporal, and channel domains. Earlier works focused on RIS implementation in only one specific domain.\\
\textbf{3) Full-wave physics and EM modeling.} Our IM implementations are complemented by realistic, full-wave physical and EM modeling of RIS. We introduce the circuit design of a unit-cell that is later used for BER analyses.\\
\textbf{4) BER analyses.} We employ our novel EM-compliant design for BER analyses of R-MTS-based SM and OFDM-IM systems. Our comparison with traditional modulation techniques under Rayleigh and Rician channels shows the feasibility of our RIS implementation. 

Throughout this paper, we denote the vector and matrices in bold lowercase and uppercase letters, respectively. The $i$-th element of a vector $\mathbf{x}$ is $\mathbf{x}_i$. The $(l,k)$-th element of a matrix $\mathbf{X}$ is $\mathbf{X}_{l,k}$. The identity matrix of size $N\times N$ is $\mathbf{I}_N$. The notations $(\cdot)^T$ and $(\cdot)^H$ denote transpose and conjugate transpose operations, respectively. The notation $\Re(\cdot)$ denotes the real part of the signal. The notation $\mathcal{C N}(\mu, \sigma^2)$ denotes complex normal distribution with mean $\mu$ and variance $\sigma$. The statistical expectation operator is $\mathbb{E}[\cdot]$. The operator $\textrm{\textrm{mod}}(\cdot)$ denotes the modulo operation.

The rest of the paper is organized as follows. In the next section, we summarize the system models used for RIS-aided IM systems. We follow this by describing our RIS system implementations of various IM techniques in Section~\ref{sec:IM-techniques}. We present our EM-compliant circuit design for the RIS meta-atom to achieve these systems in Section~\ref{sec:emAnalysis}. We validate our models and methods through numerical experiments in Section~\ref{sec:num_exp} and conclude in Section~\ref{sec:summ}.

\section{System Model}
\label{sec:sysmod}
\begin{figure}[t]
  \centering
  \includegraphics[width=85mm]{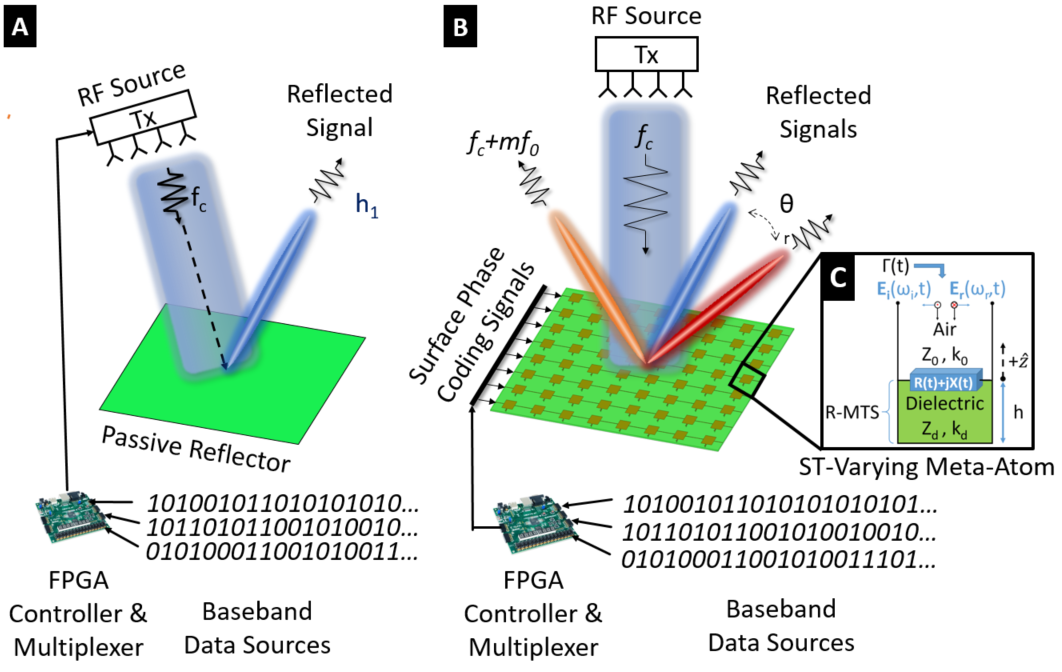}
  \caption{(a) Simplified illustration of (a) conventional $M$-ary QAM/PSK transmitter using a passive reflector and (b) our digitally programmable time-varying R-MTS transceiver comprising meta-atoms, each of which is implemented using (c) active element circuit (see Section~\ref{sec:emAnalysis}). The reflected radiation pattern scan angle is denoted as $\theta$ and the harmonic frequency of the reflected signal is $f_{c} + m f_{0}$ where $f_{c}$ is the carrier frequency, $f_{0}$ is the harmonic spacing, and $m$ is an integer value of the harmonic. The RF chain of the conventional transmitter is not shown in (a) for simplicity.}
  \label{fig:MTS_ConceptualGraphic_revA}
\end{figure}

In a conventional transmitter (Figure~\ref{fig:MTS_ConceptualGraphic_revA}a), the information bits at the baseband are modulated and multiplexed by a digital controller implemented in a field-programmable gate array (FPGA) device. After conversion to an analog signal, an RF source feeds the signal to a passive reflector which transmits the beam in the desired direction. On the other hand, our R-MTS operates in a reflect-array configuration (Figure~\ref{fig:MTS_ConceptualGraphic_revA}b), wherein the RF source is placed above the surface and excites a 2-D $M \times N$ array of sub-wavelength meta-atoms that form the RIS. The data bits are modulated onto the carrier wave through phase coding of the R-MTS. Each meta-atom consists of a resonant scattering particle with an active embedded element, such as a varactor diode \cite{hodge2018utilizing}, which programmably adjusts the reflection phase of each individual meta-atom by varying its complex impedance $Z_0$. The properties of the reflected carrier wave such as its amplitude, phase, frequency, and polarization state are changed by varying $Z_0$ of the R-MTS over space and time. Figure~\ref{fig:MTS_ConceptualGraphic_revA}(c) shows the circuit diagram of our space-time (ST)-varying meta-atom (unit cell) that we describe later in Section~\ref{sec:emAnalysis}. 

Consider a single-user MIMO system with $N_{t}$ transmit and $N_r$ receive sub-apertures. At each time index, a codeword $\mathbf{x} \in \mathbb{C}^{N_{t} \times 1}$ is generated based on the input information bits. The codeword contains the complex-valued modulated symbols. The Tx source illuminates the R-MTS to perform modulation, multiplexing, and beamforming. The resulting codeword is transmitted through the R-MTS Tx array over a narrowband statistical channel, where the delay spread is much lower than the reciprocal of the bandwidth. The baseband, discrete-time $N_{r} \times 1$ complex received signal vector is
\begin{equation}
    \mathbf{y} = \sqrt{\rho}\mathbf{H}\mathbf{x}+\mathbf{n},
\end{equation}
where $\mathbf{H} \in \mathbb{C}^{N_{r} \times N_{t}}$ is the channel matrix, $\rho$ is average received power, and $\mathbf{n}  \in \mathbb{C}^{N_{r} \times 1}$ is the additive white Gaussian noise (AWGN) with $\mathbf{n} \sim \mathcal{CN}(\mathbf{0}, \sigma_n^2 \mathbf{I}_{N_r})$. 

\begin{figure}[t]
  \centering
  \includegraphics[width=85mm]{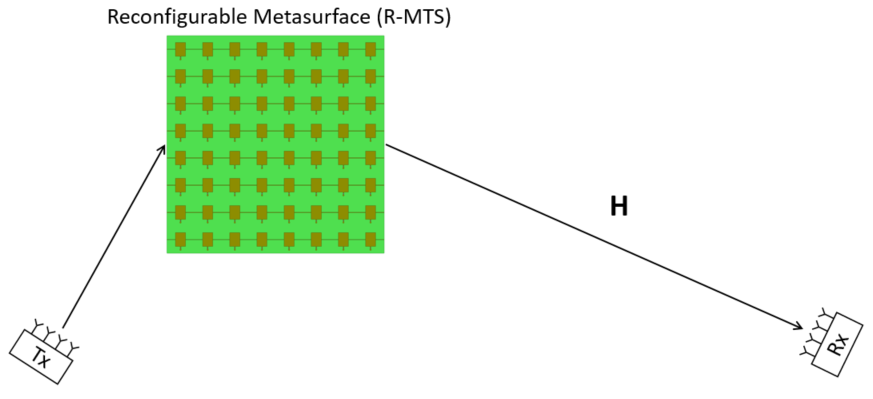}
  \caption{System model illustration of the R-MTS (RIS) transceiver for wireless communications. The Tx feed array illuminates the R-MTS with a carrier wave of frequency $f_{c}$. In this paper, we consider the R-MTS transceiver design operating through the R-MTS-to-Rx antenna channel ($\mathbf{H}$).}
  \label{fig:LIS_SystemModelFig_revF}
\end{figure}

\subsection{Channel Models}
We analyze the communications system performance of the R-MTS transceiver using several different channel models to represent specific relevant propagation environments, i.e., direct line-of-sight (LoS) channel, Rayleigh fading, and Rician fading.\\
\\\\
\textit{Direct LoS path loss}: The direct LoS component accounts for the signal path from the Tx antenna to the Rx antenna without any reflection. The direct link channel between Tx and Rx is noted as $\mathbf{h}_{TR,m}$ (Figure~\ref{fig:LIS_SystemModelFig_revF}). The channel gain of the direct LoS path is \cite{hu2020reconfigurable},
\begin{equation}
\mathbf{H}_{LoS} = \frac{\lambda}{4 \pi} \cdot \frac{\sqrt{G_{T, \operatorname{LoS}} G_{R, \operatorname{LoS}}} \cdot e^{-j 2 \pi d_{\operatorname{LoS}} / \lambda}}{d_{\operatorname{LoS}}}, \label{eqn:channel-H-los}
\end{equation}
where $\lambda$ is the wavelength of the Tx signal, $G_{T, \operatorname{LoS}}$ ($G_{R, \operatorname{LoS}}$) is the gain of the transmitter (receiver) for the direct LoS component, and $d_{\operatorname{LoS}}$ is the LoS distance between the Tx and Rx antennas.\\\\
\textit{Rayleigh fading channel}: When operating in a congested urban environment, the LoS path between the R-MTS transmitter and the intended receiver is often blocked or obstructed. The Rayleigh fading model is used for stochastic channel fading for non-LoS (nLoS) path \cite{proakis2008digital}. This model assumes that the magnitude of a signal that has passed through the channel varies randomly as per the Rayleigh distribution. 
We denote this channel by $\mathbf{H}_{nLoS} \in C^{N_{r} \times N_{t}}$, whose elements are i.i.d. complex Gaussian distributed with each element having zero mean and unit variance.\\\\
\textit{Rician fading channel}: For line-of-sight (LoS) communications, the standard statistical model for a multipath fading channel follows a Rician distribution \cite{proakis2008digital}. In Rician fading, the channel impulse response between each transmit antenna and receive antenna is modeled as the sum of the fixed LOS component and a random multipath (nLoS) channel component as \cite{paulraj2003introduction}
\begin{equation}
    \mathbf{H}_{rician} = \sqrt{\frac{K}{K+1}} \mathbf{H}_{LoS} + \sqrt{\frac{1}{K+1}} \mathbf{H}_{nLoS},
\end{equation}
where $K$ is the Rician $K$-factor of the channel, $\mathbf{H}_{LoS} \in C^{N_{r} \times N_{t}}$ is the LoS channel component, which remains unchanged within channel coherence time, and $\mathbf{H}_{nLoS} \in C^{N_{r} \times N_{t}}$ is the nLoS fading component representing random multipath fading. The Rician $K$-factor is the ratio between the power in the direct path (LoS) and the power in the other, scattered, nLoS paths. This model has been used in literature to analyze SM \cite{mesleh2008spatial} and LIS \cite{han2019large} wireless systems. 
\\

\subsection{Receiver Processing}
Assuming that the additive noise terms at the $N_{R}$ receive antennas are statistically independent zero-mean Gaussian, the maximum likelihood detector (MLD) is optimal in the sense that it minimizes the probability of error $P_{E}$. The MLD estimates the symbol vector $\hat{\mathbf{x}}$ that minimizes the Euclidean distance metric \cite{proakis2008digital}
\begin{equation}
    r(\mathbf{x}) = \sum_{l=1}^{N_{r}} {\Big| \mathbf{y}_{l} - \sum_{i=1}^{N_{t}} \mathbf{H}_{l,i}\mathbf{x}_{i} \Big|}^{2}.
\end{equation}

In particular, the signal received at the $l$-th receive antenna from the RIS is \cite{basar2020reconfigurable}
\begin{equation}
    \mathbf{y}_l = \Big[\sum_{i=1}^{N} \mathbf{H}_{l,i}e^{j\phi_{i}}\Big]\sqrt{E_s} + \mathbf{n}_{l},\ l \in \{1,...,N_{r}\},
\end{equation}
where 
$E_s = \mathbb{E}[|\mathbf{x}|^{2}]$ is the transmitted signal energy; $\mathbf{H}_{l,i}=\beta_{l,i}e^{-j\psi_{l,i}}$ is the wireless fading channel between the $l$-th receive antenna and the $i$-th meta-atom reflector element; $\beta_{l,i}$ is the corresponding complex pathloss coefficient; and $n_{l} \sim \mathcal{C N}(0, N_0)$ is the AWGN sample at the $l$-th receiver. The reflection phase of each meta-atom $\{\phi_{i}\}_{i}^{N}$ is adjusted by tuning the active embedded element to achieve the desired spatially-varying aperture phase distribution of the R-MTS. The R-MTS adjusts the reflection elements of each meta-atom to achieve beamforming, signal modulation, and spatial multiplexing. As shown in \cite{basar2020reconfigurable}, the received instantaneous SNR at the $l$-th receive antenna is 
\begin{equation}
    \gamma_{l}=\frac{\Big| \sum_{i=1}^{N} \beta_{l,i} e^{j(\phi_{i}-\psi_{l,i})} \Big|^{2} E_{s}}{N_{0}}.
\end{equation}
The SNR is maximized at the $m$-th receive antenna when the reflection phase of each meta-atom is adjusted such that $\psi_i=\psi_{m,i}$. This results in a maximum SNR at the selected receive antenna as 
\begin{equation}
    \gamma_{l}=\frac{\Big| \sum_{i=1}^{N} \beta_{m,i} \Big|^{2} E_{s}}{N_{0}}.
\end{equation}

As a performance metric, we consider the ergodic capacity of the R-MTS transceiver MIMO system. Assuming that the channel is unknown at the transmitter and known at the receiver, the instantaneous capacity of the MIMO system is \cite{paulraj2003introduction}
\begin{equation}
    C_{\mathrm{MIMO_{inst}}} = \log_{2}\Big( \det\Big( \mathbf{I}_{N_{r}} + \frac{\gamma}{N_{t}} \mathbf{H} \mathbf{H}^{H} \Big) \Big) \ \mathrm{[bit/s/Hz]},
\end{equation}
where $\mathbf{I}_{N_{r}}$ denotes the $N_{r} \times N_{r}$ identity matrix, and $\gamma$ is the average signal-to-noise-ratio (SNR) over all receiver array elements. The ergodic capacity is the expected value of the instantaneous capacity $C_{\mathrm{erg}} = \mathbb{E}_{\mathrm{\mathbf{H}}}[C_{\mathrm{inst}}]$.

\section{RIS-Aided IM Designs}
\label{sec:IM-techniques}
\begin{figure*}[t]
  \centering
  \includegraphics[width=0.99\textwidth]{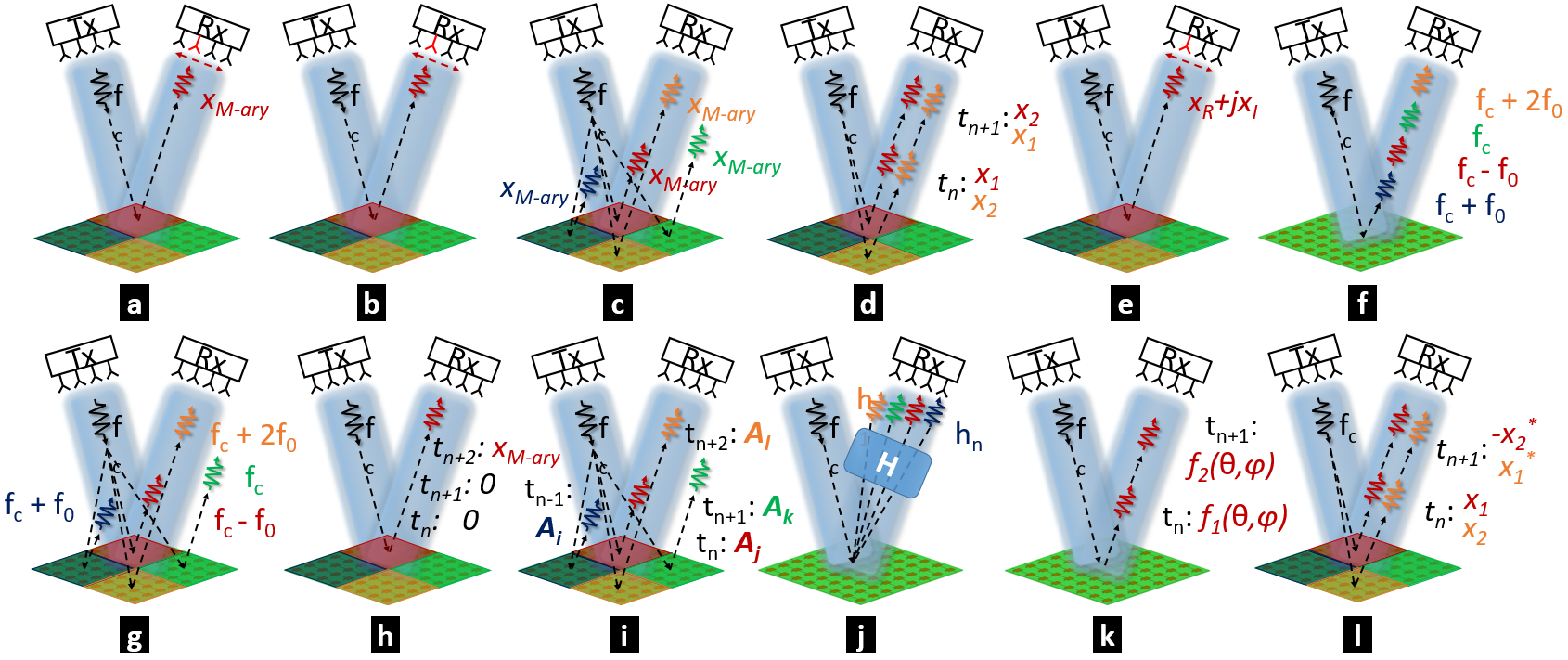}
  \caption{Selected index modulation functionalities of a ST-varying R-MTS transceiver for 6G wireless communications: (a) Classical SM, (b) SSK, (c) GSM, (d) DSM, (e) QSM, (f) SIM-OFDM, (g) MIMO-OFDM-IM, (h) SC-IM, (i) STSK, (j) MBM, (k) RA-SSK, (l) STCM.}
  \label{fig:MTS_ConceptualGraphic_panel_revA}
\end{figure*}

Table~\ref{tab:IM-types-table} lists various IM methods for which we now present RIS-aided designs. A simplified illustration of these modulations is shown in Figure~\ref{fig:MTS_ConceptualGraphic_panel_revA}.
\begin{table}[t]
\centering
\caption{Index Modulation Techniques Suitable for RISs/R-MTSs
}
\label{tab:IM-types-table}
\begin{tabular}{c|l}
\hline
\textbf{IM Domain} & \textbf{Examples} \\ [0.5ex] \hline\hline
\textbf{Phase}                               & Phase Shift Keying \cite{hodge2019reconfigurable,tang2019programmable} \\ \hline
\multirow{4}{*}{\textbf{Spatial}}   & Classical SM \cite{mesleh2008spatial}                   \\ 
                                    & Spatial Shift Keying (SSK) \cite{jeganathan2009space}   \\ 
                                    & Generalized SM (GSM) \cite{younis2010generalised} \\ 
                                    & Enhanced SM (ESM) \cite{cheng2015enhanced}     \\ 
                                    & Differential SM (DSM) \cite{bian2014differential}  \\ 
                                    & Quadrature SM (QSM) \cite{mesleh2014quadrature}  \\ \hline
\multirow{3}{*}{\textbf{Frequency}} & Subcarrier-IM On-Off Keying (SIM-OOK) \cite{hodge2019reconfigurable,zhao2018programmable,wu2019serrodyne}\\
                                    & SIM-OFDM \cite{abu2009subcarrier}     \\ 
                                    & MIMO-OFDM-IM \cite{hodge2020intelligent}    \\ \hline
\multirow{2}{*}{\textbf{Time}}      & Single-Carrier-Based (SC-IM) \cite{nakao2016single}       \\ 
\textbf{}                           & Space-Time Shift Keying (STSK) \cite{sugiura2010coherent, sugiura2011generalized}   \\ \hline
\multirow{2}{*}{\textbf{Channel}}   & Media-Based Modulation (MBM)    \cite{khandani2013media,naresh2016media,basar2019media}     \\ 
\textbf{}                           & Reconfigurable Antenna (RA-SSK) \cite{bouida2015reconfigurable,basar2017index}     \\ 
\textbf{}                           & Space-Time Channel Modulation (STCM) \cite{basar2017space,basar2017index} \\ \hline \hline
\end{tabular}
\end{table}

\subsection{Direct-Modulation Phase Shift Keying (PSK)}
The R-MTS has the ability to alter the reflection phase of each meta-atom that comprises the RIS in a programmable time-varying manner. This feature allows R-MTSs to perform $M$-ary PSK modulation without conventional RF transceiver and phase shifting circuitry. R-MTSs enable a new simplified wireless communications architecture where the R-MTS directly modulates the baseband digital signal onto the carrier wave \cite{hodge2019reconfigurable,mishra2019reconfigurable}. In this simplified architecture, the phase modulation process is accomplished by changing the time-varying reflection coefficient by applying different bias voltages to the varactor diodes embedded in each meta-atom comprising the RIS \cite{dai2019wireless}. Recently, a programmable MTS has been demonstrated as an 8-PSK wireless transmitter using this method \cite{tang2019programmable}. By dividing the metasurface into multiple radiating $n_{t}$ sub-apertures, $n_{t} \times n_{r}$ MIMO data transmission is achieved by the R-MTS as proposed in \cite{hodge2019reconfigurable} and experimentally demonstrated in \cite{tang2020wireless}.

The simplified architecture RIS-based approach to PSK eliminates the need for RF mixers, filters, and wideband power amplifiers (PAs), which increase the cost and complexity of systems. The time-varying reflection coefficient of each meta-atom in the R-MTS is expressed as
\begin{equation}
    \Gamma(t) = A(t) \times e^{j[ 2\pi f(t) \times t + \phi(t) ]},
\end{equation}
where $A(t)$, $f(t)$, and $\phi(t)$ are the amplitude, frequency, and phase of the meta-atom's reflection coefficient. The signal constellation for $M$-ary PSK is 
\begin{equation}
    \Gamma_{m} \in M = A e^{j\phi_{m}} = A e^{j\big[\frac{2(m-1)\pi}{M}\big]}, \ m = 0,1,...,M-1.
\end{equation}
The total number of bits being conveyed by the transmitter during each signaling interval in PSK is
\begin{equation}
    R_{PSK} = \log_{2}(M),
\end{equation}
where $M$ is the size of the considered PSK signal constellation.

\subsection{Spatial Modulation}
SM is a spatial domain form of index modulation that emerged as a low-complexity yet energy-efficient MIMO transmission technique \cite{mesleh2008spatial,di2014spatial,yang2014design}. Benefits of SM-MIMO for wireless communications include high throughput, power efficiency, simple RF transceiver, flexible structure, and free of inter-antenna interference (IAI) and inter-antenna synchronization (IAS). In SM, indices of activated transmit antennas convey additional bits of information in an implicit manner, leading to greater energy efficiency (EE) \cite{di2014spatial}. Communication bits are conveyed by both the indices of transmit antennas (or sub-apertures) and \textit{M}-ary constellation symbols in SM. The total number of bits being conveyed by the transmitter during each signaling interval is
\begin{equation}
    R_{SM} = \log_{2}(n_{T}) + \log_{2}(M),
\end{equation}
where $M$ is the size of the considered signal constellation, such as $M$-ary phase shift keying ($M$-PSK) or $M$-ary quadrature amplitude modulation ($M$-QAM). 

\begin{figure}[t]
  \centering
  \includegraphics[width=85mm]{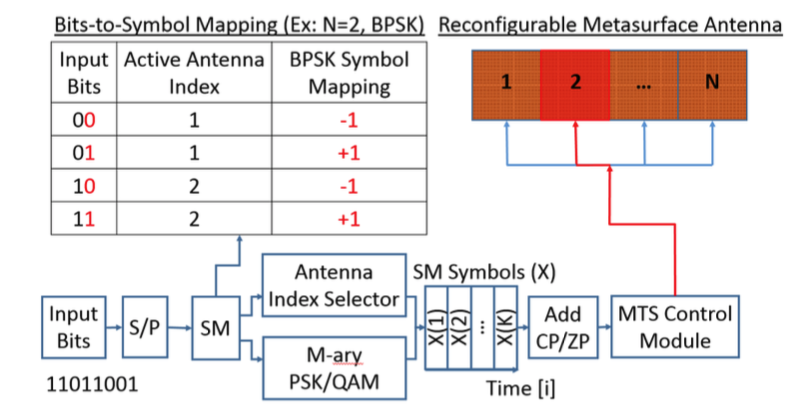}
  \caption{Simplified system block diagram of an active R-MTS transceiver-based transmit chain \cite{hodge2019reconfigurable}.}
  \label{fig:SM_R-MTS_revA}
\end{figure}

SM is particularly well suited for R-MTS/RIS implementation due to its simplified RF transceiver architecture and ability to use only a single RF source \cite{hodge2019reconfigurable}. Figure~\ref{fig:SM_R-MTS_revA} shows RIS-based SM transmit system.

Other variations of SM implemented directly using a RIS (R-MTS) transceiver include spatial shift keying (SSK) \cite{jeganathan2009space}, generalized SM (GSM) \cite{younis2010generalised}, enhanced SM (ESM) \cite{cheng2015enhanced}, differential SM (DSM) \cite{bian2014differential}, and quadrature SM (QSM) \cite{mesleh2014quadrature}. Here, SSK is the most simple form of SM where information bits are only carried by the index of a single activated Tx antenna (or sub-array of an R-MTS) without amplitude/phase modulation (APM). In GSM, multiple Tx antennas or R-MTS sub-arrays are activated to transmit the same symbol for diversity gain over the channel. The ESM is implemented with high-flexibility of antenna activation patterns and APM constellation alphabets \cite{mao2018novel}. DSM is a differently encoded space-time shift keying (DSTSK) modulation scheme that does not require accurate estimation of the channel state information (CSI) \cite{cheng2018index}. QSM is a variation where SM is conducted on the I/Q components of the signal individually to increase the number of bits transmitted. The number of bits transmitted through QSM at each time instant is $2\log_{2}(n_{T}) + \log_{2}(M)$ \cite{basar2017index}.

\subsection{Frequency IM}
We consider the following two forms of IM that employ frequency as an index.
\subsubsection{Subcarrier-Index Modulation On-Off Keying (SIM-OOK)} \label{sec:SIM-OOK}
In addition to RIS/LIS-based spatial IM, a time-modulated RIS also functions as a low-complexity and low-cost frequency-domain IM (FD-IM) transceiver. 
Space-time (ST) modulation of the RIS reflection phase enables tunable frequency translation and programmable frequency harmonics \cite{hodge2019reconfigurable,zhao2018programmable,wu2019serrodyne}. Through the time-varying phase modulation sequence of the meta-atom, the phase of the generated reflected frequency harmonic is electronically controlled. Using the beamforming capabilities of the R-MTS, these reflected frequency harmonics are steered to the direction of the receive antenna or in a way that minimizes CSI. In the most simple instantiation, data bits of the reflected carrier signal can be conveyed by the index of the frequency subcarrier. By phase modulating the frequency harmonic generated from the R-MTS, $M$-ary PSK or QAM can be used to transmit additional bits on a single frequency subcarrier. In this manner, the total number of bits being conveyed during each signaling frame interval is 
\begin{equation}
    R_{\mathrm{SIM-OOK}} = \log_{2}(n_{\mathrm{SIM}}) + \log_{2}(M),
\end{equation}
where $n_{\mathrm{SIM}}$ is the number of different discrete frequency possible harmonics that can be generated by the ST-modulated R-MTS. Section~\ref{sec:emAnalysis} explains the EM theory and analysis of R-MTS frequency harmonic and subcarrier generation.

\subsubsection{Subcarrier-Index Modulation OFDM (SIM-OFDM)}

\begin{figure*}[t]
  \centering
  \includegraphics[width=1.0\textwidth]{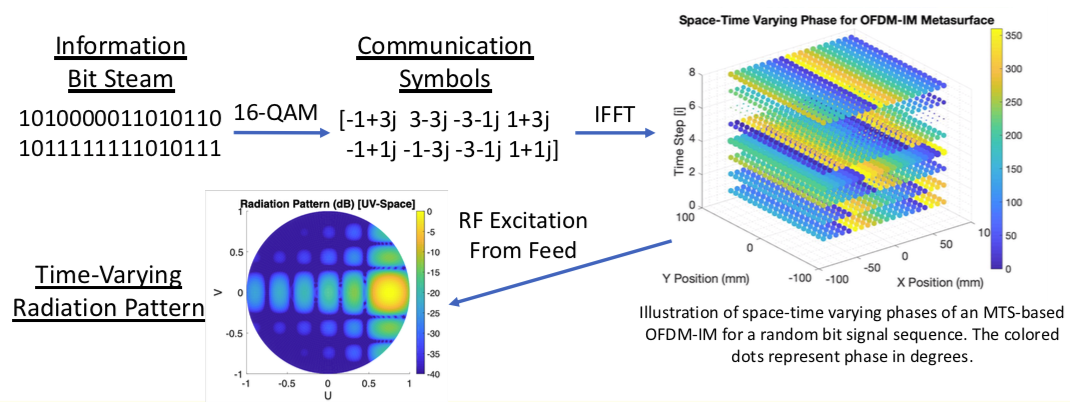}
  \caption{Illustration of the R-MTS transceiver implementation of OFDM-IM (SIM). The input data bit stream is converted into $16$-QAM communication symbols inside of the R-MTS's digital controller. The OFDM-IM block creator in the R-MTS's digital controller then uses an inverse fast Fourier transform (IFFT) to frequency multiplex communication symbols. Rather than using every available subcarrier index for bit transmission, this FD-IM technique uses the permutation of subcarrier-index OOK to convey information bits to the receiver. The R-MTS digital controller generates a space-time varying phase and amplitude matrix. Time-varying voltage control signals are used to tune the diodes embedded in each meta-atom of the R-MTS to have the desired complex reflection coefficient. Colored dots represent the phase of each meta-atom in degrees. The size of each dot indicates its amplitude. Through spatial and temporal modulation of the R-MTS's EM surface response, the R-MTS transceiver directly modulates the carrier wave and transmits the OFDM-IM into the desired spatial direction to maximize SNR at the receiver.}
  \label{fig:MTS_ConceptualGraphic_ofdmim_revA}
\end{figure*}

A common type of digital modulation used in current wireless communications, such as IEEE 802.11a wireless local area networks (LAN), WiMAX, and 4G LTE, is OFDM \cite{proakis2008digital}. Here, multiple closely spaced orthogonal frequency subcarrier signals with overlapping spectra are transmitted to carry data in parallel. A generic OFDM transmitter takes the form shown in Figure~\ref{fig:MTS_ConceptualGraphic_ofdmim_revA}. 
If $N_{\mathrm{sub}}$ frequency carriers are used, and each subcarrier is modulated using a $M$-ary signal constellation, the OFDM symbol alphabet consists of $M^{N_{\mathrm{sub}}}$ combined symbols. Advantages of OFDM include high SE, robustness against ISI and multi-path, and efficient FFT implementation. However, OFDM is also sensitive to Doppler-shift induced ICI and suffers from a high peak-to-average-power ratio (PAPR), requiring linear transmitter circuitry, which results in poor power efficiency \cite{cheng2018index}. The traditional OFDM transmit signal is 
\begin{align}
    x(t) = \frac{1}{\sqrt{N_{\mathrm{sub}}}} \sum_{a=0}^{N_{\mathrm{sub}}-1}X_{a}e^{j2\pi at/T_{sym}}, 0 \leq t \leq T_{sym},
\end{align}
where $X_{a}$ are the information symbols and $T_{sym}$ is the temporal symbol duration. When the complex symbol is up-converted for wireless transmission, the transmitted signal is 
\begin{align}
    s(t) = \Re(x(t)e^{j2\pi f_{c}t})
\end{align}
where $f_{c}$ is the carrier frequency of the transmitter.

The currently employed modulation techniques based on MIMO-OFDM are not sufficient to satisfy the requirements for 5G and upcoming 6G networks \cite{cheng2018index}. Conventional MIMO may achieve high SE with a large number of antennas. However, it suffers from EE limitations because of the linearly increasing power consumption of a large number of RF chains. To overcome these SE and EE challenges, index modulation has been proposed as an advanced novel modulation technique \cite{basar2020reconfigurable}. Similar to SM in the spatial domain, index modulation is applied to the frequency domain using SIM-OFDM \cite{abu2009subcarrier}, also known as OFDM-IM. In SIM, activated subcarriers are determined by the corresponding majority bit-values of an OOK data stream \cite{mao2018novel}. SIM (OFDM-IM) has also been combined with MIMO to form a MIMO-OFDM-IM transmission scheme that can achieve significantly better error performance than traditional MIMO-OFDM \cite{hodge2020intelligent}.

To overcome physical layer challenges associated with MIMO-OFDM, SIM-OFDM, and MIMO-OFDM-IM, we propose the use of a space-time varying RIS (R-MTS) transceiver to implement these frequency domain (FD)-IM techniques. Through space-time coding of an R-MTS, frequency harmonics or subcarriers are generated and spatially steered in a controllable manner to produce SIM \cite{hodge2019reconfigurable}. For R-MTS implementation of OFDM and OFDM-IM, signal amplitude and phase control can each be controlled either in the RF feed circuitry or directly through switching in the R-MTS. Generation of controllable harmonics using time-varying R-MTS enables direct implementation of SIM, OFDM-IM, FSK, and other FD modulations.


\subsection{Time slot IM}
Time-domain (TD) IM schemes can be classified as either IM carried out on time slots or IM carried out on dispersion matrices \cite{mao2018novel}. This gives rise to the following RIS-aided implementations.
\subsubsection{Single-Carrier-Based IM (SC-IM)}
 SC-IM is a TD-IM scheme that performs IM on time slots of each transmit frame, where only part of the time slots are occupied for data transmission \cite{nakao2016single}. In SC-IM, index bits are conveyed by the indices of activated time slots with no additional energy consumption. As a result, SC-IM results in enhanced EE compared to traditional SC Frequency Domain Equalization (SC-FDE) techniques. An important benefit of SC-IM is its reduced peak-to-average power ratio (PAPR) compared to OFDM-IM and conventional OFDM \cite{ishikawa201850}. At the receiver, the data symbols and corresponding index bits are jointly estimated using the MLD. Assuming that $k$ time slots are activated in each sub-frame, the data rate of SC-IM is \cite{mao2018novel}
\begin{equation}
    R_{\mathrm{SC-IM}} = \frac{N_{s} \Big( k \log_{2}M + \Big[ \log_{2} \left(^{l_{s}}_{k}\right) \Big]\Big)}{(N_{s} + L_{CP})l_{s}}  \ \mathrm{[bit/frame]},
\end{equation}
where $N_{s}$ is the number of symbols per frame, $l_{s}$ is the number of time slots in each sub-frame. $L_{CP}$ is the length of the cyclic prefix (CP), and $M$ is the size of the $M$-ary symbol constellation.

Using our software-defined R-MTS transceiver, SC-IM can be implemented by letting the digital controller map the data into index bits and $M$-ary constellation bits. Similar to the R-MTS implementation of SM, the index bits are mapped from the controller into transmission time frames and the R-MTS directly modulates the phase and/or amplitude of the carrier signal at the appropriate time slot to modulate the $M$-ary constellation bits.

\subsubsection{Space-Time Shift Keying (STSK)}
The STSK is a modulation scheme for MIMO communication systems,  where the concept of SM is extended to include both the space and time dimensions \cite{sugiura2010coherent}. Instead of using time slots, TD-IM is carried out in a different manner on dispersion matrices. In STSK, index bits are conveyed by the single activated dispersion matrix \cite{mao2018novel}. Unlike SM, STSK has the benefit of being capable of striking a flexible trade-off between diversity gain and multiplexing gain. In STSK, for each signaling block, one out of the $Q$ available dispersion matrices is selected for data transmission where the index of the dispersion matrix selected carries additional implicit information bits. To enhance the achievable data rate, generalized STSK (GSTSK) has been developed by activating $P>1$ dispersion matrices simultaneously \cite{sugiura2011generalized}. However, GSTSK comes at the cost of increased detection complexity. STSK is a special case of GSTSK where $P=1$. The achievable data rate for GSTSK is \cite{mao2018novel}
\begin{equation}
    R_{\mathrm{GSTSK}} = \frac{\log_{2} \left(^{Q}_{P}\right) + P\log_{2}M}{N_{ns}}  \ \mathrm{[bpcu]},
\end{equation}
where $N_{ns}$ is the number of time slots per signaling interval and $\mathrm{bpcu}$ is \textit{bits per channel use}, which is a measure of spectral efficiency.

A low-cost ST-modulated R-MTS transceiver is well suited to be the transmitter of a STSK or GSTSK system. The digital controller of the R-MTS can map input data bits into dispersion matrix index bits and $M$-ary constellation bits. The digital controller of the R-MTS transceiver can also select the $P$ dispersion matrices from the set of $Q$ pre-designed dispersion matrices
\begin{equation}
    \mathcal{A} = \{ \mathbf{A}_{1}, \mathbf{A}_{2}, ... , \mathbf{A}_{Q} \},
\end{equation}
where $\mathbf{A}_{i} \in C^{N_{t} \times N_{ns}}$ are complex valued matrices. Through ST-modulation, the R-MTS is also able to perform the ST-mapper functionality of the GSTSK transmitter. Similar to our R-MTS-based implementation of SM, the R-MTS can also implement the more general SM by choosing the appropriate transmit sub-aperture or element during the specified time slot.

\subsection{Channel IM}
Finally, the channel or paths may be employed for index modulation as follows.
\subsubsection{RA-SSK} Reconfigurable antennas (RAs) have long been a subject of EM research to the wireless communications community \cite{christodoulou2012reconfigurable}. Today, RAs are necessary in many modern telecommunication systems and wireless networks. An RA is an antenna capable of modifying its frequency and radiation properties dynamically and in a controlled manner \cite{bernhard2007reconfigurable}. These radiation properties also include the antenna's polarization state. To provide a dynamic response, RAs typically integrate an inner active element, such as RF switches, diodes, mechanical actuators, or tunable materials, that enable the intentional redistribution of the RF currents over the antenna surface in a controllable manner. By performing multiple functions with a single antenna, RAs often provide the benefit of reduced SWaP-C and the amount of hardware required in systems. The software-defined functionality of R-MTSs/RISs makes these low-cost apertures ideal platforms for implementing dynamic radiation patterns with strict radiation requirements \cite{mishra2019reconfigurable}.

Recently, RA-based IM modulation techniques have also become a focus of wireless communication research \cite{basar2020reconfigurable}. Specifically, RA-SSK has been introduced as a modulation technique for MIMO systems operating over Rician fading channels \cite{bouida2015reconfigurable}. RA-SSK has been shown to provide better BER performance and reduced system complexity compared to conventional SSK in an LOS radio propagation channel. In RA-SSK, the Tx antenna's radiation pattern state is used exclusively to encode information. Assuming a multipath RF propagation environment, the received signal depends on the RF interaction between the spatial distribution of scatterers in the channel and the radiation pattern of the antenna. By altering the radiation pattern of the RA, the SM concept is extended to encode index bits of information in the interplay between the antenna's radiation pattern and nearby scatterers in the channel \cite{basar2020reconfigurable}. In \cite{bouida2015reconfigurable}, BER performance is enhanced by altering the polarization state of the antenna to minimize the effective Rician $K$ factor of the channel.

The challenge for RA-SSK systems is to minimize the spatial correlation between different radiation pattern states \cite{bouida2015reconfigurable}. The throughput (bits per signaling interval) of a RA-SSK transmitter is
\begin{equation}
    R_{\mathrm{RA-SSK}} = \log_{2}N_{t} + \frac{1}{N_{t}} \sum_{i}^{N_{t}} \log_{2}N_{s_{i}},
\end{equation}
where $N_{s_{i}}$ is the number of uncorrelated radiation pattern states available at the $i$-th transmit antenna or R-MTS sub-array. Practically, the number of index bits that are signaled is limited by the number of distinct (uncorrelated) radiation patterns generated by the RA. As we show later, the degrees of freedom and a large number of tunable elements provided by an R-MTS/RIS have the potential to increase the number of available uncorrelated radiation pattern states and improve the BER performance of RA-SSK systems.

\subsubsection{Media-Based Modulation (MBM)}
Media-based modulation (MBM) is a form of channel-domain IM (CD-IM) that is implemented by intentionally altering the far-field radiation pattern of a reconfigurable antenna (RA) \cite{hodge2020media}. In MBM, an RA is used to perturb the propagation environment to create independent channel fade realizations which themselves are used as the constellation points. Hypothetical RF mirrors have been proposed to implement MBM by creating different channel fade realizations based on the ON/OFF status of each RF mirror \cite{naresh2016media}. This study showed that for the same spectral efficiency, GSM-MBM achieves better performance compared to MIMO-MBM. However, they did not present a physically realizable implementation or EM models of the RF mirrors necessary to implement MBM. 
\begin{figure*}
\begin{center}
\noindent
  \includegraphics[width=1.0\textwidth]{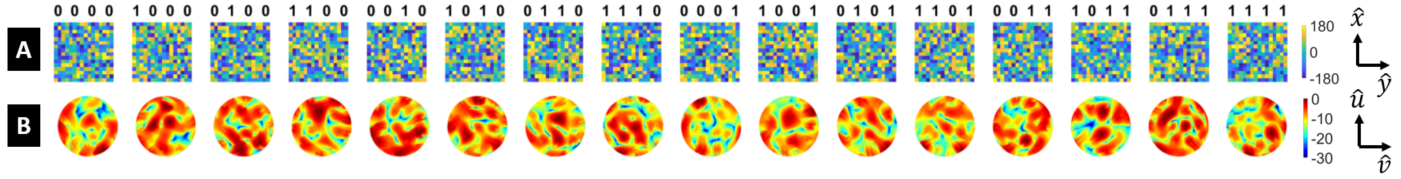}
  \caption{Generation of 16 distinct radiation patterns to represent the signal constellation by varying the reflection phase coding of the R-MTS that can be used for MBM transmission of 4 bpcu. (a) 16 pseudo-random R-MTS surface reflection phase codings (in degrees). (b) Normalized radiation patterns (in dB) calculated from each respective R-MTS phase coding in (a). Each radiation pattern activated a different channel state response through its interaction with scatterers in the propagation environment. The normalized radiation patterns are plotted using the UV-coordinate system with U and V spanning from -1 to 1 representing the half-space above the R-MTS.\vspace{-14pt}}\label{fig:Fig1Label3}
\end{center}
\end{figure*}

Very recently, an RA simulation model using a traditional 2-by-2 square patch array consisting of four microstrip patch antennas was presented to provide four different radiation pattern states for MBM \cite{hodge2020media}. This RA implementation is feasible but has limited degrees of freedom (DoF) to optimize the transmitted radiation patterns for particular channel states. Here, novel RA architectures that generate a sufficiently high number of antenna states with relatively low correlation are needed for the successful implementation of MBM. The R-MTSs are an ideal physical layer antenna technology to realize RF mirrors. By controlling the reflection phase of each meta-atom, the R-MTS is able to transmit a diverse set of tightly tailored and customized radiation patterns Figure~\ref{fig:Fig1Label3}. The R-MTS transceiver antenna provides $P^{N}$ of potential radiation pattern states, where $P$ is the number of tunable phase states of each meta-atom and $N$ is the number of meta-atom elements in the R-MTS. As shown in Section~\ref{sec:r-mtsBeamforming}, the R-MTS transceiver provides the radiation pattern reconfigurability of a traditional phased array without the RF complexity, RF losses, and cost associated with mixers phase shifter and phase shifters \cite{hodge2019reconfigurable,tang2020wireless}. This R-MTS is also applicable for other MBM techniques such as GSM-MBM or layered MIMO-MBM (LMIMO-MBM) \cite{mao2018novel}.

\subsubsection{Space-Time Channel Modulation (STCM)}
As an alternative to MBM and RA-SSK, STCM has been proposed as a channel-based IM technique that combines RAs and space-time block coding (STBC) \cite{hodge2019reconfigurable}. In STCM, index bits are conveyed by the indices of selected channel state vectors in Alamouti STBC systems \cite{hodge2019reconfigurable,mao2018novel}. Using an R-MTS transceiver, the STBCs are applied to the transmit signal in the R-MTS's digital controller.

\section{EM Design}
\label{sec:emAnalysis}

The behavior an electromagnetic medium can generally be expressed using Maxwell's equations as 
\begin{align}
    \textbf{D}=\epsilon_{0}\textbf{E} + \textbf{P}_{e}, \\
    \textbf{B}=\mu_{0}\textbf{H} + \textbf{P}_{m},
\end{align}
where $\epsilon_{0} = 8.854 \cdot 10^{-12}$ As/Vm and $\mu_{0} = 4\pi \cdot 10^{-7}$ are the free-space permittivity and permeability, $\textbf{E}$ (V/m) and $\textbf{H}$ (A/m) are the electric and magnetic fields, $\textbf{D}$ (As/m$^2$) and $\textbf{B}$ (Vs/m$^2$) are the electric and magnetic flux density fields, and $\textbf{P}_{e}$ (As/m$^2$) and $\textbf{P}_{m}$ (Vs/m$^2$) are the electric and magnetic polarization densities of the medium. We consider $\textbf{E}$ as the field excitation for the surfaces and media that form the MTS. In particular, metasurfaces that exhibit magnetoelectric coupling are called bianisotropic metasurfaces \cite{asadchy2018bianisotropic}. The EM surface properties of space-time varying MTS can be modeled in terms of equivalent sheet impedance/admittance, susceptibility, or polarizability. The EM susceptibilities of an electrically thin MTS are modeled as
\begin{align}
    \begin{pmatrix}
        \textbf{P}_{e} \\
        \textbf{P}_{m}
    \end{pmatrix}
    =
    \begin{pmatrix}
        \epsilon_{0}\overline{\overline\chi}_{ee} & \sqrt{\epsilon_{0}\mu_{0}} \: \overline{\overline\chi}_{em} \\
        \sqrt{\epsilon_{0}\mu_{0}} \: \overline{\overline\chi}_{me} & \mu_{0}\overline{\overline\chi}_{mm}
    \end{pmatrix}
    \cdot
    \begin{pmatrix}
        \textbf{E} \\
        \textbf{H}
    \end{pmatrix}
    =
    \overline{\overline\chi} \cdot
    \begin{pmatrix}
        \textbf{E} \\
        \textbf{H}
    \end{pmatrix}
    ,
\end{align}
where $\overline{\overline\chi}_{ee}$, $\overline{\overline\chi}_{em}$, $\overline{\overline\chi}_{me}$, and $\overline{\overline\chi}_{mm}$ are 3 x 3 dyadic tensors representing the electric-to-electric, electric-to-magnetic, magnetic-to-electric, and magnetic-to-magnetic susceptibility terms and $\overline{\overline\chi}$ represents the dyadic 6 x 6 tensor that combines the previous four susceptibility coupling terms. To achieve communications modulation, multiplexing, and antenna beamforming, we consider space-time varying R-MTSs. The EM responses of each meta-atom in the R-MTS vary by time, spatial frequency $k$, and temporal frequency $\omega$ frequency such that $\overline{\overline\chi}$(t, $k$, $\omega$). To avoid unwanted fast-time dispersion effects, we only consider meta-atom time variations $\Delta t >> T=\frac{2\pi}{\omega}$ \cite{caloz2019spacetime}.

\begin{figure}[t]
  \centering
  \includegraphics[width=0.8\columnwidth]{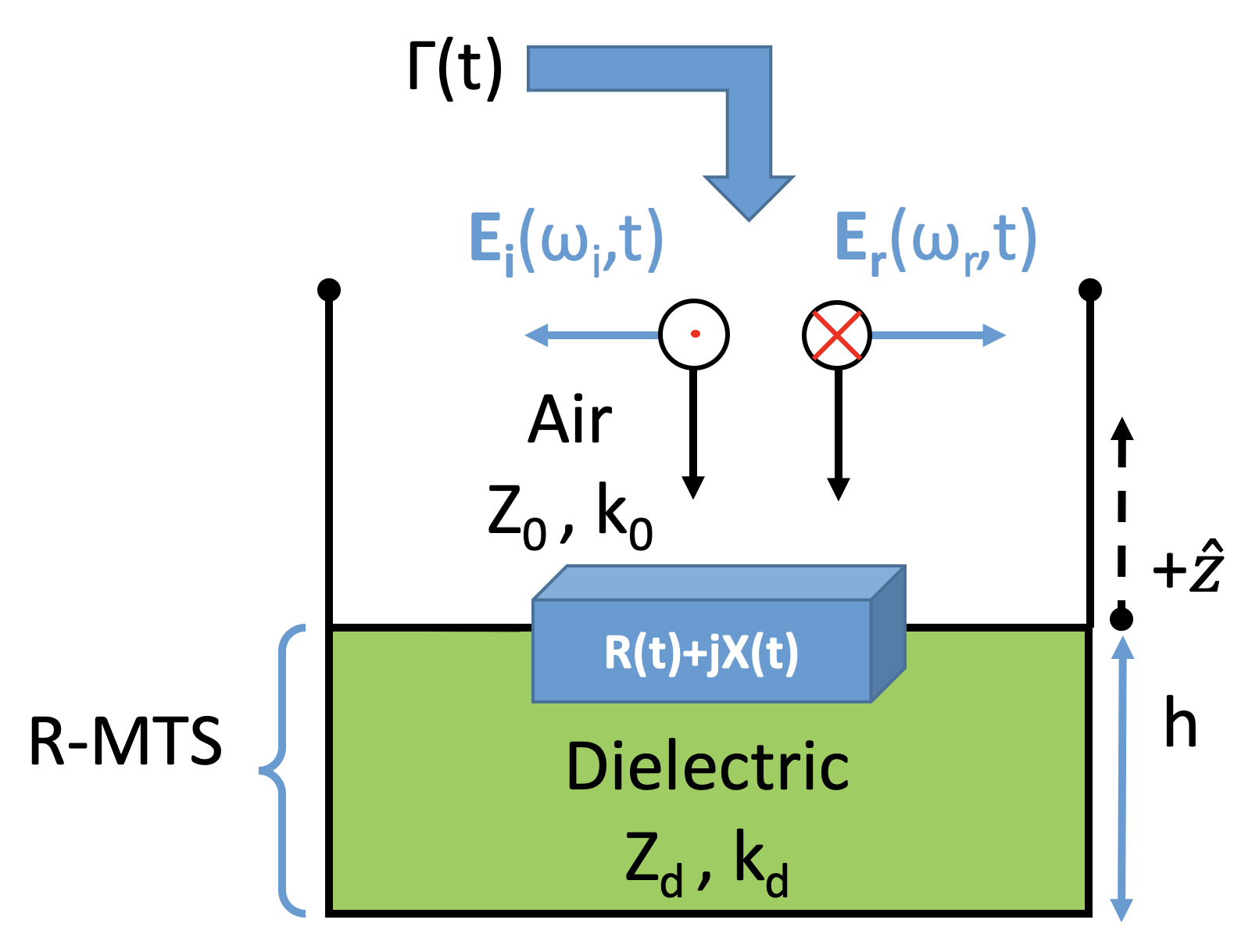}
  \caption{Equivalent circuit model representation of the R-MTS meta-atom. The complex impedance of the MTS is represented by $Z(p,q,t) = 1 / {Y(p,q,t)} = R(t) + jX(t)$.}
  \label{fig:equivCircuit}
\end{figure}

Consider a planar R-MTS in a reflect-array configuration, consisting of a dense periodic array of sub-wavelength elements, whose response is programmable and varied in time. Figure \ref{fig:equivCircuit} depicts the R-MTS and its equivalent circuit. Given that the meta-atom unit-cell dimensions are much smaller than the operating wavelength, the MTS is characterized by its effective surface admittance $Y_{s}$ \cite{glybovski2016metasurfaces}. In general, time-modulated MTSs exhibit a frequency dispersive surface admittance $Y_{s}=Y_{s}(\omega,t)$. However, if the modulation-induced frequency shift is much smaller than the operating frequency, modulation-induced dispersion effects are neglected and surface admittance is only considered as a function of time $Y_{s}=Y_{s}(t)$ \cite{ramaccia2019phase}.

Since our reflect-array MTS is fully reflective with no transmitted fields, our analysis only considers the incident and reflected EM fields above the MTS ($z>0$). The total EM fields are a superposition of the incident and reflected fields as 
\begin{align}
    \boldsymbol{E}_{i}(z,t) = \Re[\boldsymbol{E}_{0,i}e^{j(\omega_{i}t-k_{i}z)}], \label{eqn:defineFields1} \\
    \boldsymbol{E}_{r}(z,t) = \Re[\boldsymbol{E}_{0,r}e^{j(\omega_{r}t+k_{r}z)}], 
    \label{eqn:defineFields2}
\end{align}\normalsize
where $\boldsymbol{E}_{0,i(r)}$ are the incident and reflected complex electric field amplitudes, $k_{i(r)}$ are the wavenumbers, $\omega_{i(r)}$ are the respective angular frequencies, and $\Re[\cdot]$ denotes the real-part of a complex quantity. At the surface of the R-MTS $(z=0)$, the incident and reflected electric fields are related by the time-varying complex reflection coefficient $\Gamma(t)$:
\begin{align}
    \boldsymbol{E}_{r}(z=0,t) = \Gamma(t) \boldsymbol{E}_{i}(z=0,t).  \label{eqn:A4}
\end{align}\normalsize

Knowing the incident E-field from our RF source, the reflection coefficient at each meta-atom in the R-MTS can be time-modulated to achieve a reflected wave with the desired frequency, amplitude, and phase at each time step. From (\ref{eqn:defineFields1}-\ref{eqn:A4}), the time-varying reflection coefficient from an incident wave frequency $\omega_{m}$ is \cite{ramaccia2019phase}
\begin{align}
    \Gamma(\omega_{i},t) = \frac{\boldsymbol{E}_{r}(z=0,t)}{\boldsymbol{E}_{i}(z=0,t)} = \frac{\boldsymbol{E}_{0,r}}{\boldsymbol{E}_{0,i}}e^{j(\omega_{r}-\omega_{i})t}. \label{eqn:A2B}
\end{align}

The meta-atom unit cell is modeled using the equivalent circuit model shown in Figure \ref{fig:equivCircuit}. The time-varying complex reflection coefficient is related to the surface admittance of the R-MTS as
\begin{align}
    \Gamma(t) = \frac{Y_{0}-Y_{s}(t)}{Y_{0}+Y_{s}(t)},
    \label{eqn:A2}
\end{align}
where $Y_{0}=1/120\pi$ is the free-space admittance. By electrically controlling the time-varying reflection phase of each meta-atom, any arbitrary phase profile can be formed across the R-MTS array. Assuming a full 360-degree range of meta-atom phase variation, each point along the R-MTS exhibits properties resembling a perfect-electric conductor (PEC), perfect magnetic conductor (PMC), or any reflection phase in between. 

\subsection{R-MTS Beamforming} \label{sec:r-mtsBeamforming}
The active MTS is in a reflectarray configuration \cite{mishra2019reconfigurable} and controlled by an MTS control module, which is an FPGA controller that calculates the necessary phase of each unit cell to perform MTS beam scanning and phase modulation. The MTS control module also sends voltage signals to each unit cell to realize the desired phase states. 

Assume that the R-MTS is illuminated by a plane wave with time-harmonic dependence ($e^{j2\pi f_{c}t}$) at normal incidence ($\theta_{i}=0$). The reflected far-field radiation pattern of the R-MTS is approximated as \cite{yang2016programmable,wan2016field,zhang2018space}
\begin{align}
\begin{split}
    &f(\theta,\phi,t) = \sum_{p=1}^{M}\sum_{q=1}^{N}E_{pq}(\theta,\phi)\Gamma_{pq}(t) \\
    &\exp \Big\{ j\frac{2\pi}{\lambda_{c}} \big[(p-1)d_{x}\sin\theta\cos\phi + (q-1)d_{y}\sin\theta\sin\phi \big] \Big\}, \label{eqn:ff-rad-pattern}
    \end{split}
\end{align}
where $\theta$ and $\phi$ are the elevation and azimuth angles of an arbitrary direction, $E_{pq}(\theta,\phi)$ is the far-field pattern of the $(p,q)$-th meta-atom element computed at the carrier frequency $f_{c}$, the time-varying reflection coefficient of the $(p,q)$-th meta-atom of the R-MTS is $\Gamma_{pq}(t) = a_{e}(p,q,t) \exp{(j \varphi_{e}(p,q,t))}$, $d_x$ and $d_y$ are the meta-atom unit cell grid spacing in the $x$ and $y$ dimensions, and the wavenumber is $\lambda_{c} = \frac{1}{f_{c}}$. The directivity of the R-MTS antenna aperture in terms of the far-field radiation pattern \eqref{eqn:ff-rad-pattern} scattered by the metasurface is 
\begin{equation}
    \mathrm{Dir}(\theta,\phi,t) = \frac{4\pi {|f(\theta,\phi,t)|}^{2} }{\int_{0}^{2\pi} \int_{0}^{\pi/2} {|f(\theta,\phi,t)|}^{2} \sin\theta d\theta d\phi}. \label{eqn:ff-pattern-dir}
\end{equation}

Next, we synthesize a formulation to determine the phase coding of the R-MTS to steer the reflected radiation pattern to an arbitrary spatial direction of interest. Assume $\theta_r$ and $\theta_i$ are the reflected and incident wave directions, respectively. If $\frac{d\Phi}{dx}$ is the gradient of phase discontinuity along the MTS, then the MTS beam scanning is modeled using the generalized law of reflection as \cite{yu2011light}
\par\noindent\small
 \begin{equation}
 \sin(\theta_r)-\sin(\theta_i) = \frac{\lambda_0}{2\pi}\frac{d\Phi}{dx}.
 \end{equation}\normalsize
 Assuming $\theta_i=0$ and taking MTS unit cell parameters into account, the beam scanning equation becomes \par\noindent\small
\begin{equation}
\theta_r = \sin^{-1}{\left(\frac{\lambda_0}{2\pi}\frac{\Delta\phi_{e}}{Qd}\right)
\label{eqn:beam_scan2}
},
\end{equation}\normalsize
where $\Delta\phi_{e}$ is the phase tuning range of the unit cell element, $Q$ is the selected period of unit cells, and $d$ is the element spacing of each unit cell. From this relation, we calculate the phase difference between each unit cell necessary to scan the beam in the desired direction. The FPGA-based MTS control module computes the values of $\Delta\phi_{e}$ and $Q$ accordingly for beam scanning and calculates the phase of each unit cell column ($\phi_{\mathrm{steer}}(p)$) using \par\noindent\small 
\begin{align}  
\phi_{\mathrm{steer}}(p) &= \textrm{mod}{(p/Q)}(\Delta\phi_{e}/Q).\label{eqn:f4}
\end{align}\normalsize
The unit cell grid spacing in each dimension is denoted by $d_x$ and $d_y$ and the wavenumber is $k_c = \frac{2\pi}{\lambda_c}$. The far-field phase of each element due to the array factor of the MTS is \par\noindent\small
\begin{align} 
\phi_{ff}(p,q) &= k_c \sin{\theta} ((p-1)d_x\cos{\phi} + (q-1)d_y\sin{\phi}).\label{eqn:f3}
\end{align}\normalsize

Denote $\phi_{\textrm{mod}}(p,q,t)$ as the time-varying phase modulation term that accounts for the desired phase shift keying (PSK) or quadrature amplitude modulation (QAM) modulation symbols. The reflection amplitude and phase of each MTS unit cell are $a_e(p,q)$ and $\phi_e(p,q)$, respectively \cite{wan2016field}. The desired reflection phase of each unit cell is \par\noindent\small
\begin{align} 
\phi_{e}(p,q,t) &= \phi_{\textrm{mod}}(p,q,t) +\phi_{\mathrm{steer}}(p,q) + \phi_{ff}(p,q).\label{eqn:f2}
\end{align}\normalsize

From (\ref{eqn:f4})-(\ref{eqn:f2}), the far-field radiation pattern of the signal-modulating-MTS is\par\noindent\small
\begin{align} 
 f(\phi,\theta,t) &= \sum_{m=1}^{M} \sum_{n=1}^{N}a_{e}(p,q)\exp{\{j\phi_{e}(p,q,t)\}}.\label{eqn:f1}
\end{align}\normalsize


\subsection{R-MTS Harmonic Generation} \label{sec:R-MTS-Harmonic-Generation}

To analyze the response of the time-varying R-MTS, we take the Fourier transform (FT) of \eqref{eqn:A4} using convolution theorem \cite{ramaccia2019phase} to obtain
\begin{align}
    \boldsymbol{E}_{r}(\omega) = \Gamma(\omega)*\boldsymbol{E}_{i}(\omega) = \int\Gamma(\omega-\omega')\boldsymbol{E}_{i}(\omega')d\omega',  \label{eqn:A5}
\end{align}\normalsize
 where $\omega$ represents the frequency spectrum and $*$ denotes the convolution operation. It follows from \eqref{eqn:A5} that the frequency of the incident wave $\omega_{i}$ is converted to the spectrum of the reflected frequencies according to the frequency spectrum of the reflection coefficient. The time-varying reflection coefficient $\Gamma(t)$, varied by periodically modulation $Y_{s}(t) = 1/Z_{s}(t)$, is expressed as a Fourier series,
\begin{align}
    \Gamma(t) = \sum_{n}\Gamma^{n}(\omega_{i})e^{jn\omega_{m}t},  \label{eqn:A6}
\end{align}\normalsize
 where $\omega_{m}$ is the meta-atom's angular modulation frequency and $n$ represents each frequency harmonic. Taking FT of \eqref{eqn:A6} results in
\begin{align}
    \Gamma(\omega) = \sum_{n}\Gamma^{n}(\omega_{i})\delta(\omega-n\omega_{m}).  \label{eqn:A7}
\end{align}\normalsize

 The frequency spectrum of the EM field reflected from the time-modulated R-MTS is obtained from \eqref{eqn:A5} and \eqref{eqn:A7} as \cite{ramaccia2019phase}
\begin{align}
    \boldsymbol{E}_{r}(\omega) = \sum_{n}\Gamma^{n}(\omega_{i})\boldsymbol{E}_{0,i}\delta(\omega-\omega_{i}+n\omega_{m}).  \label{eqn:A8}
\end{align}\normalsize
 It follows that the signal reflected from the time-modulated R-MTS is the illuminating incident wave frequency plus and minus frequency harmonics of the meta-atom phase modulation frequency $\omega_{m}$. From \eqref{eqn:A2} and \eqref{eqn:A8}, the required time-varying surface admittance $Y_{s}(t)$ is a function of $\omega_{m}$:
\begin{align}
    Y_{s}(t) = -jY_{0}\tan(\omega_{m}t/2).  \label{eqn:A9}
\end{align}\normalsize

The Fourier series coefficients $a_{pq}^{m}$ of the periodic function $\Gamma_{pq}(t)$ are \cite{zhang2018space}
\begin{align}
    a_{pq}^{m} = \sum_{n=1}^{L}\frac{\Gamma_{pq}^{n}}{L} \sinc \Big(\frac{\pi m}{L}\Big) \exp \Big[\frac{-j\pi m (2n-1)}{L} \Big], \label{eqn:f100}
\end{align}
where $L$ is the number of time-steps in steps in the coding sequence per modulation signaling period and $m$ represents the $m$-th frequency harmonic of the carrier frequency $f_{c}$. From this equation, we can see that the amplitude and phase of the $m$-th frequency harmonic reflected from the R-MTS is a function of its time-varying reflection coefficient $\Gamma_{pq}^{n}$. Key parameters for RIS-based frequency harmonic generation are modulation period ($T_{0}$), harmonic spacing ($f_{0} = 1/T_{0}$), and harmonic frequency ($f_{c} + m f_{0}$). We show that the desired frequency subcarriers for FD-IM wireless communication techniques can be generated by a R-MTS by phase (and/or amplitude) modulation of each meta-atom. The far-field radiation pattern of the space-time varying R-MTS at the $m$-th harmonic frequency $f_{c} + m f_{0}$ is 
\begin{align}
\begin{split}
    &f_{m}(\theta,\phi) = \sum_{p=1}^{M}\sum_{q=1}^{N}E_{pq}(\theta,\phi) \: a_{pq}^{m} \\
    &\exp \Big\{ j\frac{2\pi}{\lambda_{c}} \big[(p-1)d_{x}\sin\theta\cos\phi + (q-1) d_{y} \sin\theta\sin\phi \big] \Big\}.  \label{eqn:f101}
    \end{split}
\end{align}
This shows that we have the ability to generate and spatially steer frequency harmonics in a controllable manner for advanced RIS-based modulation, multiplexing, and beamforming. The R-MTS implementation of a multi-function RIS transceiver gives us many degrees of freedom and the flexibility to optimize our wireless communication techniques for the mission and the environment.



\subsection{Design of R-MTS}

\begin{figure}[t]
  \centering
  \includegraphics[width=1.0\columnwidth]{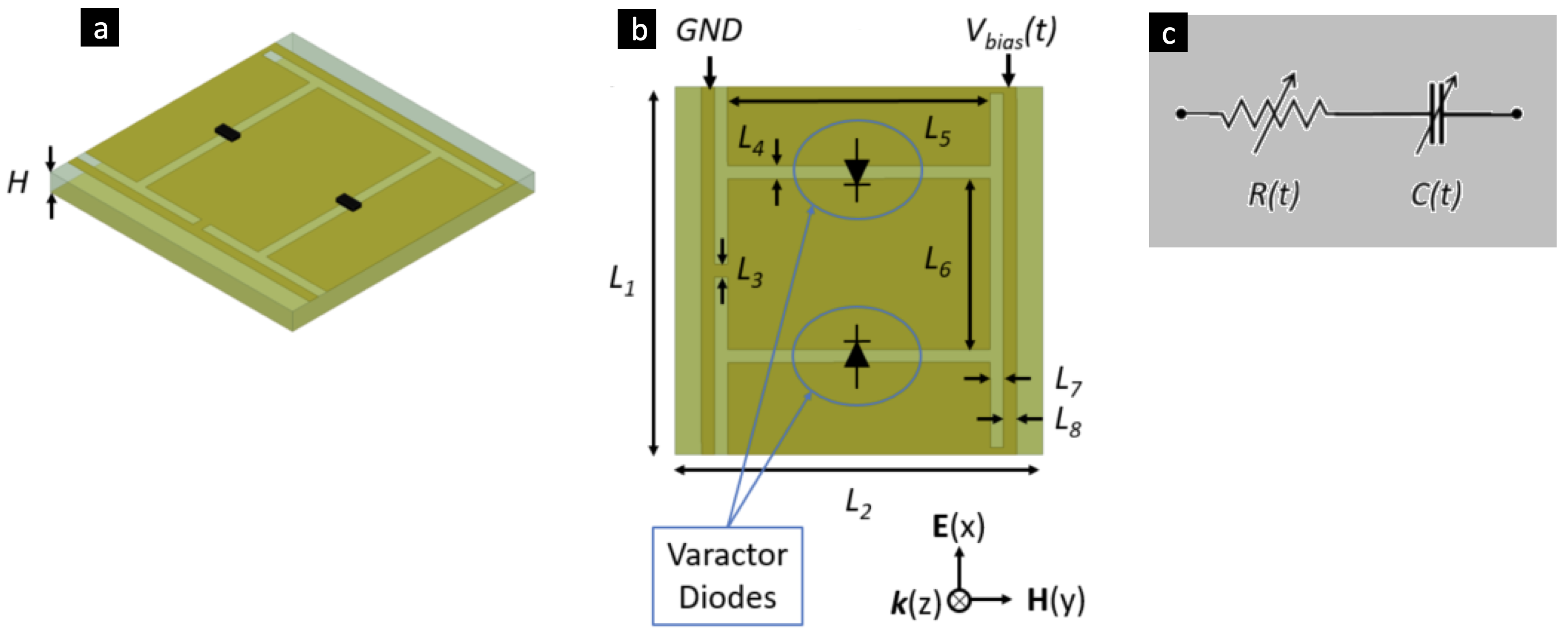}
  \caption{Illustration of the time-varying R-MTS meta-atom considered in this paper. (a) Isometric view of the unit cell. (b) Top-down view of the meta-atom unit cell. The physical dimensions of each meta-atom unit cell are: $H$ = 0.2 mm, $L_{1}$ = 2.8 mm ($\lambda_{c}/3.83$), $L_{2}$ = 2.8 mm ($\lambda_{0}/3.83$), $L_{3}$ = 0.1 mm, $L_{4}$ = 0.1 mm, $L_{5}$ = 2.0 mm, $L_{6}$ = 1.3 mm, $L_{7}$ = 0.1 mm, $L_{8}$ = 0.1 mm. The metal traces (top layer) and groundplane (bottom layer) are copper. The dielectric layer of thickness $H$ is a RT/Duroid{\textregistered} 5880 ($\epsilon_{r} = 2.2$, $\tan \delta = 0.0009 $) substrate. (c) Circuit model of our tunable varactor diode consists of a tunable series resistance ($R$) and capacitance ($C$).}
  \label{fig:unitCellDiagramAndDims_v2}
\end{figure}

We designed a reflective time-varying R-MTS (Figure~\ref{fig:unitCellDiagramAndDims_v2}) where each meta-atom unit cell is embedded with two varactor diodes. The tunable diodes enable the reflection phase of the meta-atom to be continuously tuned across a wide phase range ($>300$ deg) by changing the bias voltage to the diodes. An isometric view of the meta-atom unit cell is shown in Figure~\ref{fig:unitCellDiagramAndDims_v2}a. Many of the recent MTS designs in literature \cite{yang2016programmable,zhao2018programmable,zhang2018space,dai2019wireless} require vias in the dielectric substrate of the MTS to electrically ground the embedded tunable diodes. A meta-atom design with vias allows control signals to be routed to each individual meta-atom, however, vias also increase the cost and complexity of the fabrication process. To reduce fabrication cost and complexity of this design, we chose a meta-atom architecture that does not require vias. 

The specific meta-atom design used in this work is adapted from \cite{wu2019serrodyne}. However, our meta-atom design differs in that is modified for a reflective R-MTS operation rather than as a transmission MTS-based phase shifters. We selected the dimensions of the meta-atom to provide an operational reflection phase tuning range of $>300$ deg at $f_{c}$ = 28 GHz. Physical dimensions of R-MTS meta-atom design are shown in Figure~\ref{fig:unitCellDiagramAndDims_v2}b. We synthesized the dimensions of the meta-atom unit cell through parametric tuning and trial-and-error iteration using ANSYS HFSS full-wave EM simulation. The metal traces (top layer) and groundplane (bottom layer) of our meta-atom are copper. The dielectric layer of thickness $H$ is a RT/Duroid{\textregistered} 5880 ($\epsilon_{r} = 2.2$, $\tan \delta = 0.0009 $) substrate. The circuit model for our tunable varactor diode chip with tunable capacitance and resistance is shown in Figure~\ref{fig:unitCellDiagramAndDims_v2}c. This tunable diode representation allows each column to have a tunable and programmable reflection phase and amplitude. The tunable diode elements integrated into the R-MTS surface provide locally and continuously tunable complex surface impedance. The time-varying reflection phase is primarily controlled by varying $C$ and amplitude modulation is performed directly by the meta-atom by varying $R$.

\begin{figure}[t]
  \centering
  \includegraphics[width=80mm]{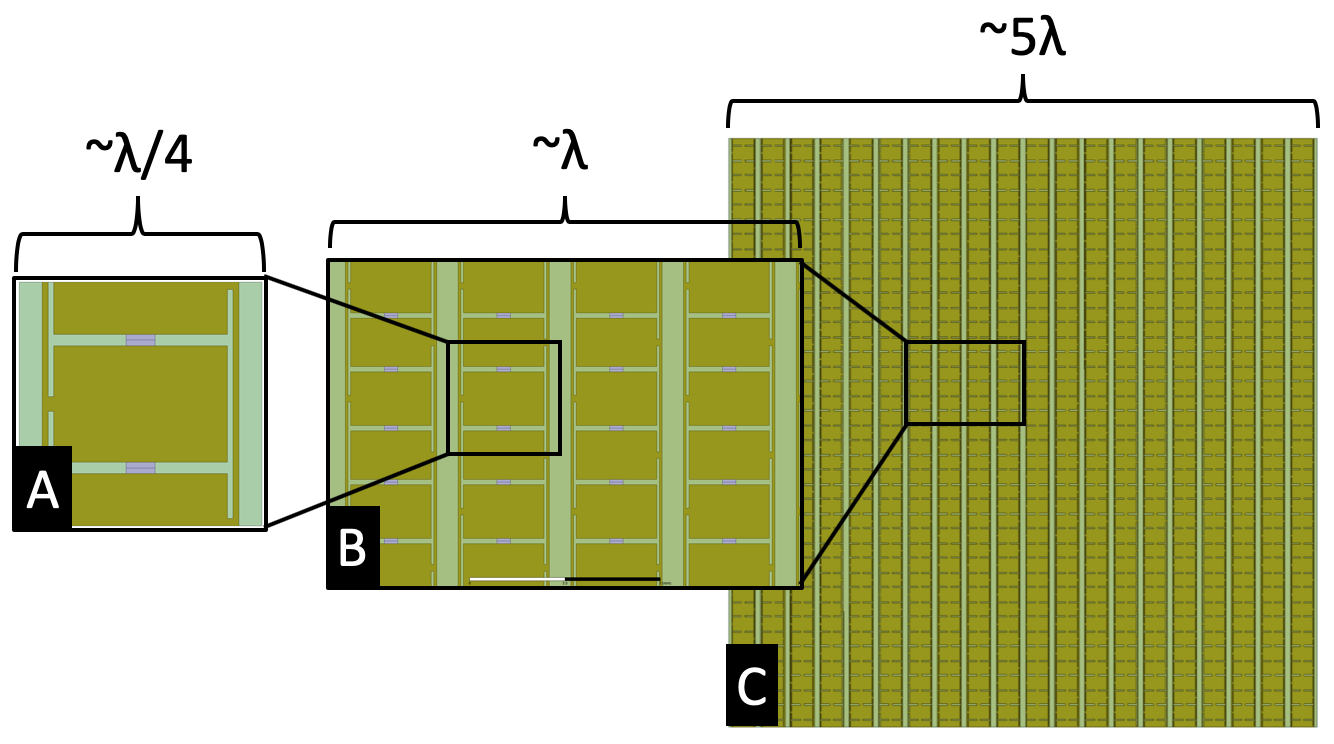}
  \caption{Multi-scale view of the time-varying R-MTS considered in this paper. (a) The meta-atom unit cell grid dimensions are approximately $\lambda/4$ and at $28$ GHz. (b) Meta-atoms are arranged in columns on a rectangular lattice. (c) The $20 \times 20$ meta-atom R-MTS considered in this paper is approximately $5\lambda \times 5\lambda$. The macroscopic R-MTS dimensions (\# of meta-atoms) can be scaled to meet system requirements.}
  \label{fig:MTS_blowupDiagram_revA}
\end{figure}

The R-MTS aperture consists of an array of meta-atoms (Figure~\ref{fig:MTS_blowupDiagram_revA}). The R-MTS can be designed to arbitrary dimensions based on the required system performance and SWaP-C constraints. In this paper, we nominally consider a $20 \times 20$ meta-atom R-MTS. The meta-atoms are arranged in columns on a rectangular lattice. Control signals independently address the top and/or bottom half of each column to spatially steer the reflected beam to the desired receiver or to create multiple sub-apertures for MIMO and spatial IM techniques. A larger R-MTS apertures provide more antenna gain for the transmitter, which results in higher SNR at the receiver, or a greater number of sub-apertures for greater communication capacity. Based on phase coding of the aperture, the R-MTS forms multiple simultaneous beams using either multiple sub-apertures or a shared full-aperture.

\section{Numerical Experiments}
\label{sec:num_exp}
We now provide physics- and EM-compliant models of our proposed RIS design. A major limitation of current research on RISs in wireless networks is the lack of accurate and tractable models that describe the R-MTS as a function of their EM properties \cite{basar2019wireless}. Several studies on RIS and LIS antenna systems assume meta-atoms to be perfect (lossless) reflectors with a full $360$ tuning range without providing EM simulation results or verification of proposed conceptual designs \cite{basar2019wireless, basar2020reconfigurable}. To realize RISs for real-world wireless communication, EM-compliant modeling and simulation of the R-MTSs RF response over frequency, angle, and polarization are needed. To help bridge this gap between the communication and EM research communities, we provide EM-compliant simulation results of our proposed IM-based R-MTS transceiver at both the meta-atom and finite R-MTS array level.

\subsection{Tuning range}
\begin{figure}[t]
  \centering
  \includegraphics[width=85mm]{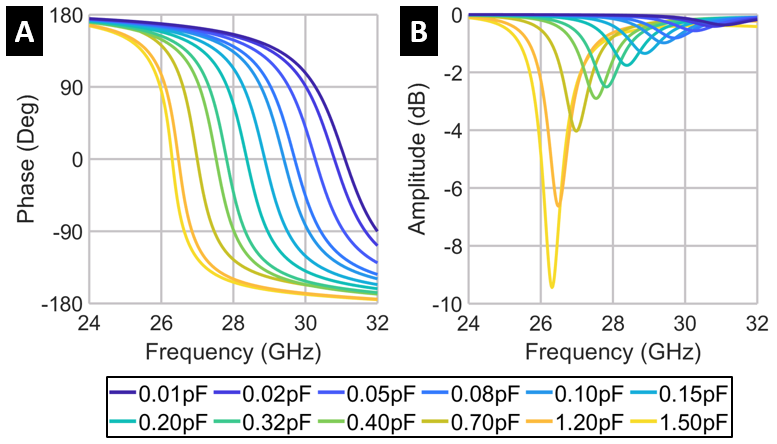}
  \caption{Simulated tunable reflection (a) phase and (b) amplitude of the R-MTS meta-atom presented in this paper, shown in Figure~\ref{fig:unitCellDiagramAndDims_v2}. Capacitance of the embedded tunable diode is varied from $C = 0.01-1.50$ pF and resistance is $R = 0.5$ Ohms. (a) At $28$ GHz, this meta-atom has a tunable reflection phase range of $310$ deg. This tunable phase range is the EM mechanism of the R-MTS that is used to modulate the communication signal for modulation techniques described in \ref{sec:IM-techniques}. (b) At $28$ GHz, the meta-atom is relatively low-loss with a reflection amplitude of $<2.2$ dB for all simulated tuning states.}
  \label{fig:tunable_phase}
\end{figure}

The simulated reflection phase and corresponding amplitude responses of the tunable meta-atom are shown in Figure~\ref{fig:tunable_phase}. By tuning the capacitance of the diodes embedded in the meta-atom from $C = 0.01-1.50$ pF, a phase tuning range of $310$ deg is achieved at $28$ GHz (Figure~\ref{fig:tunable_phase}a). At this frequency, the meta-atom is relatively low-loss with a reflection amplitude of $<2.2$ dB for all simulated tuning states (Figure~\ref{fig:tunable_phase}b). By achieving a $300+$ deg phase tuning range for the meta-atom, we are able to use nearly the full phase tuning range for $M$-ary PSK modulation and associated IM modulation techniques described in Section~\ref{sec:IM-techniques}.

\begin{figure*}
\begin{center}
  \centering
  \includegraphics[width=1.0\textwidth]{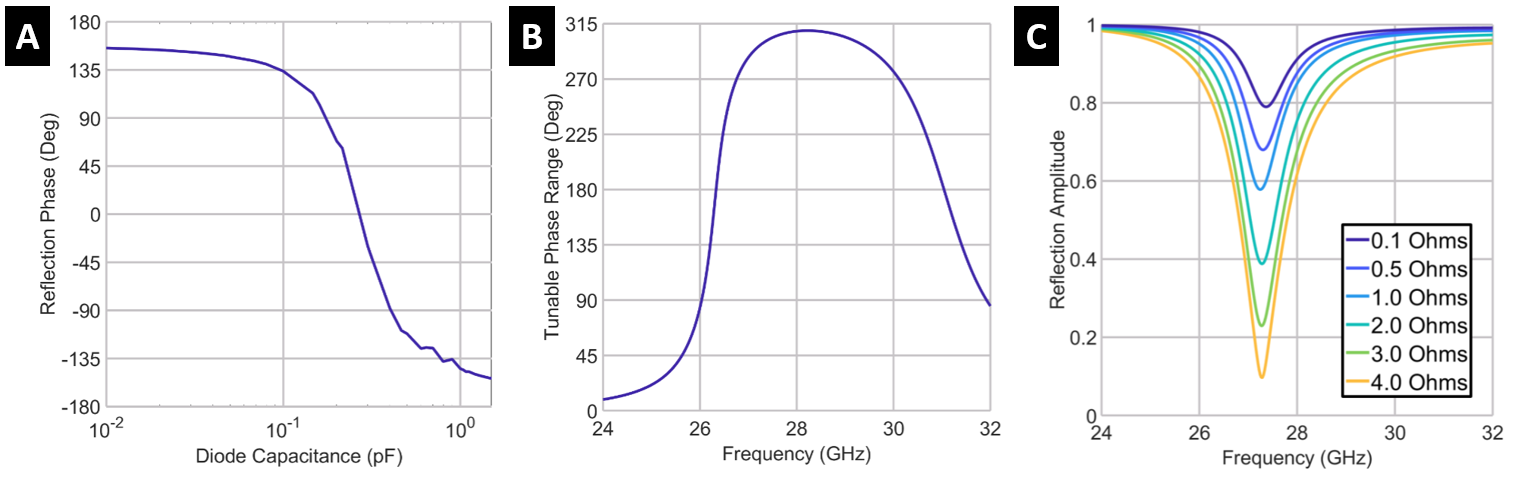}
  \caption{(a) Relationship between the tunable diode capacitance ($C$) and the reflection phase of the meta-atom at a frequency of $28$ GHz. This data is taken from the HFSS simulation results shown in Figure~\ref{fig:tunable_phase}. (b) Simulated tunable reflection phase range of the R-MTS meta-atom presented in this paper versus frequency. The maximum tunable phase range of this meta-atom is $310$ deg at $28.2$ GHz. This meta-atom has a tunable phase range of $>270$ deg from $26.8-30.1$ GHz ($12.5\%$ bandwidth) and a tunable phase range of $>180$ deg from $26.3-30.0$ GHz ($17.5\%$ bandwidth). This data is taken from the HFSS simulation results shown in Figure~\ref{fig:tunable_phase}. (c) Simulated reflected amplitude of the meta-atom while varying the tunable diode resistance from $R = 0.1-4.0$ Ohms. The amplitude of the reflected signal can be modulated for wireless communication by time-varying the resistance of the diode chip embedded in the meta-atom.}
  \label{fig:phase_vs_cap}
\end{center}
\end{figure*}

To better understand the relationship between the reflection phase and the diode capacitance, we plot the phase response from Figure~\ref{fig:tunable_phase}a at $28$ GHz as function of diode capacitance in Figure~\ref{fig:phase_vs_cap}. We observe a sharp change ($\sim270$ deg) in the reflection phase from $C = 0.1-1.0$ pF. These results indicate that a majority of the phase tuning range can be achieved in a more limited range of diode capacitance. Below $C = 0.1$ pF, the reflection phase begins to reach a saturation region. However, achieving a near-$360$ phase tuning range using a single-layered surface requires an order of magnitude of greater of addition capacitance tuning range. In practice, the phase tuning range of the meta-atom is limited by the capacitance tuning range of the selected diode. A two-layer meta-atom with an impedance inverter between them can be used achieve a near-$360$ deg phase tuning range within a practical capacitance tuning range \cite{wu2019serrodyne}. However, this approach increases SWaP-C of the R-MTS antenna by increasing required thickness, doubling the number of embedded diodes, increasing power consumption, and increasing system complexity.

The reflection phase of the meta-atom can be varied by changing the capacitance of the surface (Figs.~\ref{fig:tunable_phase}-\ref{fig:phase_vs_cap}). However, QAM-based $M$-ary signal modulation requires independent control of both amplitude and phase. To provide reflection amplitude control independent of the phase control, we include a tunable resistor in our embedded diode chip as shown in Figure~\ref{fig:unitCellDiagramAndDims_v2}. The simulation results showing the impact of varying the tunable diode resistance ($R$) on the reflection amplitude are shown in Figure~\ref{fig:phase_vs_cap}c. In this simulation, the resistance is varied from from $R = 0.1-4.0$ Ohms. At $28$ GHz, the amplitude tuning range for this range of capacitance is $0.61-0.91$. However, at $27.3$ GHz, which is within the frequency bandwidth of this meta-atom design, the tunable amplitude range is $0.1-0.79$. By using larger values of $R$, it is likely that a wider amplitude tuning range can be achieved at $28$ GHz. It should be noted that this method of increasing surface $R$ for signal amplitude modulation reduces system efficiency due to the dissipative losses of increasing $R$. A more complex, yet more energy efficient, system architecture is to modulate the amplitude of the Tx signal in the feed circuitry and directly modulate the phase of the carrier wave using the R-MTS. However, this approach also requires more RF circuitry in the feed. The goal of the R-MTS design presented in this paper is to minimize cost and complexity of the RIS/R-MTS transceiver.


\subsection{Radiative performance}
\begin{figure}[t]
  \centering
  \includegraphics[width=1.0\columnwidth]{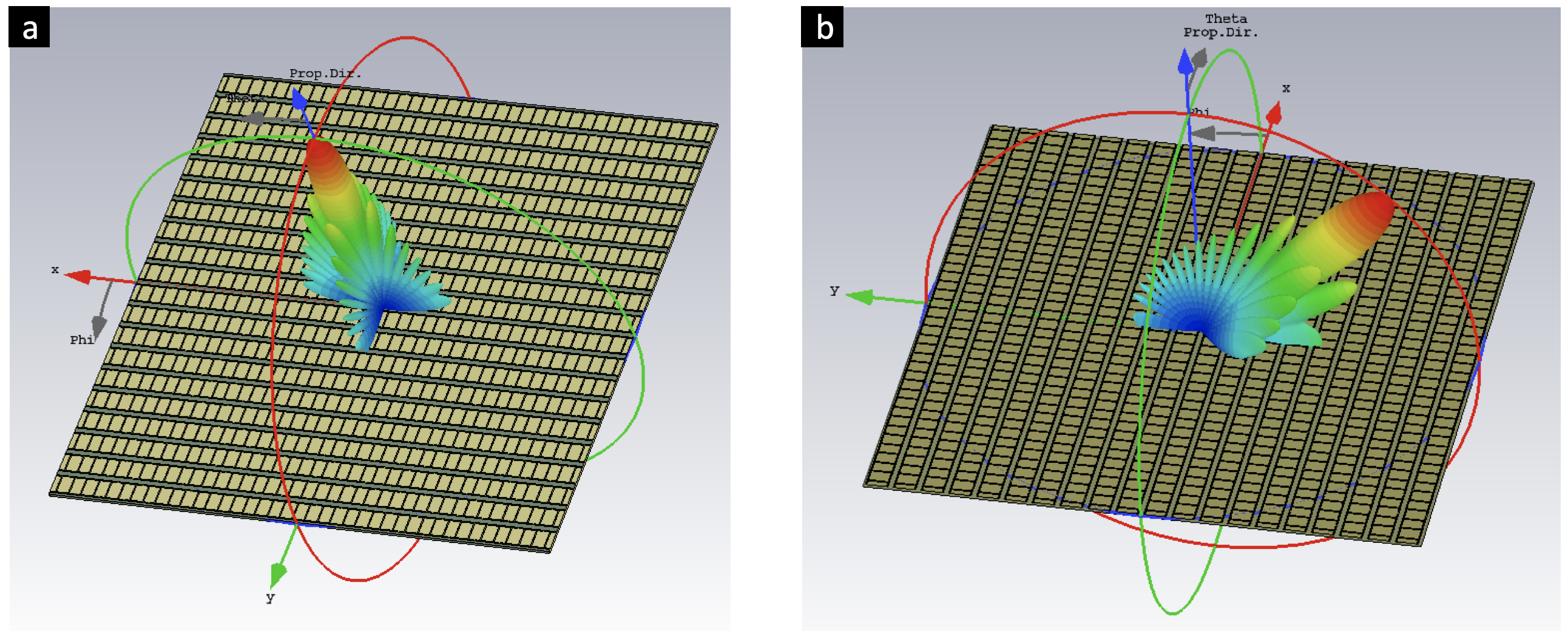}
  \caption{Finite array EM simulation of the $20 \times 20$ meta-atom R-MTS using CST Microwave Studio. The R-MTS finite array is illuminated by a linearly polarized plane wave traveling in the $-z$ direction with the E-field oriented in the $+x$ direction. a) Phase shift of each meta-atom set to uniform phase states to scan the R-MTS antenna's radiation pattern to a peak at $\theta = 0$ deg, $\phi = 0$ deg (broadside). b) The reflection phase of each tunable meta-atom column are adjusted using \eqref{eqn:f4} to steer the radiation pattern to a peak at at $\theta = 45$ deg, $\phi = 0$ deg. By adjusting the phase coding of the R-MTS, the radiation pattern can be arbitrarily steered and reconfigured using \eqref{eqn:f1}, including for the IM communication techniques described in Section~\ref{sec:IM-techniques}.}
  \label{fig:Beam_Scanning_CST_Graphic_revA_191206}
\end{figure}

To examine the radiation performance of the macroscopic R-MTS aperture (shown in Figure~\ref{fig:MTS_blowupDiagram_revA}c), we perform finite-array EM simulation analysis of the $20 \times 20$ element R-MTS using CST Microwave Studio's time-domain EM solver. The results for the R-MTS array's radiation pattern are shown in Figure~\ref{fig:Beam_Scanning_CST_Graphic_revA_191206}. The R-MTS finite array is illuminated by a linearly polarized plane wave traveling in the $-z$ direction with the E-field oriented in the $+x$ direction to simulate the RF source. By varying the phase coding of the R-MTS, we show through EM simulation that the radiation pattern of the R-MTS can be steered to a desired spatial direction according to \eqref{eqn:f4}-\eqref{eqn:f1}. As examples, we show the radiation pattern steered to $\theta = 0$ deg (Figure~\ref{fig:Beam_Scanning_CST_Graphic_revA_191206}a) and $\theta = 45$ deg (Figure~\ref{fig:Beam_Scanning_CST_Graphic_revA_191206}b). The shape of the radiation pattern behaves as predicted.

\subsection{Harmonic generation}
\begin{figure}[t]
  \centering
  \includegraphics[width=1.0\columnwidth]{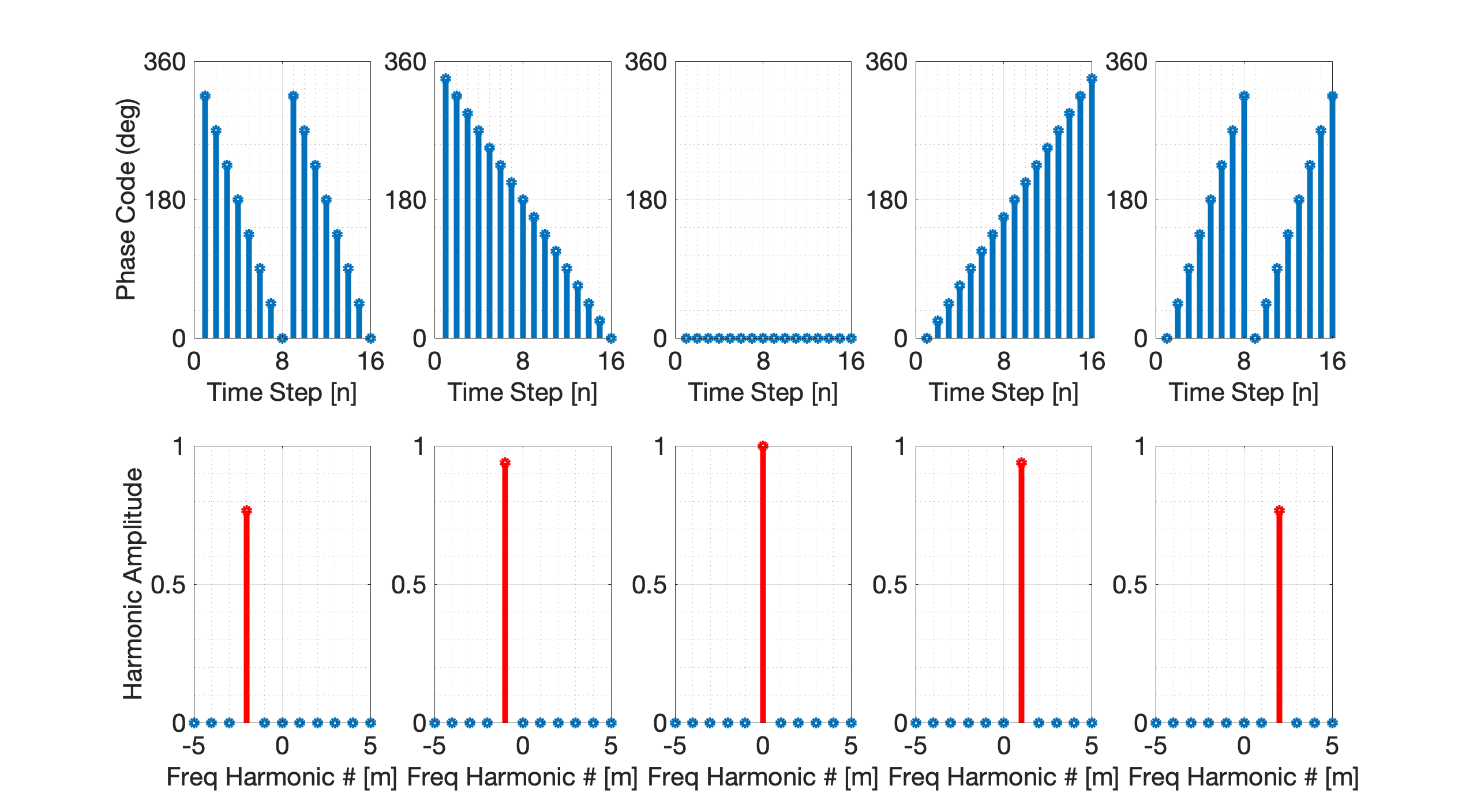}
  \caption{Numerical simulation of R-MTS harmonic generation from temporal modulation of the R-MTS's reflection phase. By controlling the phase coding sequence of each meta-atom via a time-varying bias voltage signal, the carrier wave at frequency $f_{c}$ is translated into a desired frequency harmonic $f_{c} + m f_{0}$. The slope of the meta-atom's time-modulated phase coding sequence is varied to generate single frequency harmonics from $m = -2$ to $m = +2$. The modulation period, harmonic frequency, and harmonic spacing are adjusted per the discussion in Section~\ref{sec:R-MTS-Harmonic-Generation}.}
  \label{fig:harmonic1}
\end{figure}

\begin{figure}[t]
  \centering
  \includegraphics[width=1.0\columnwidth]{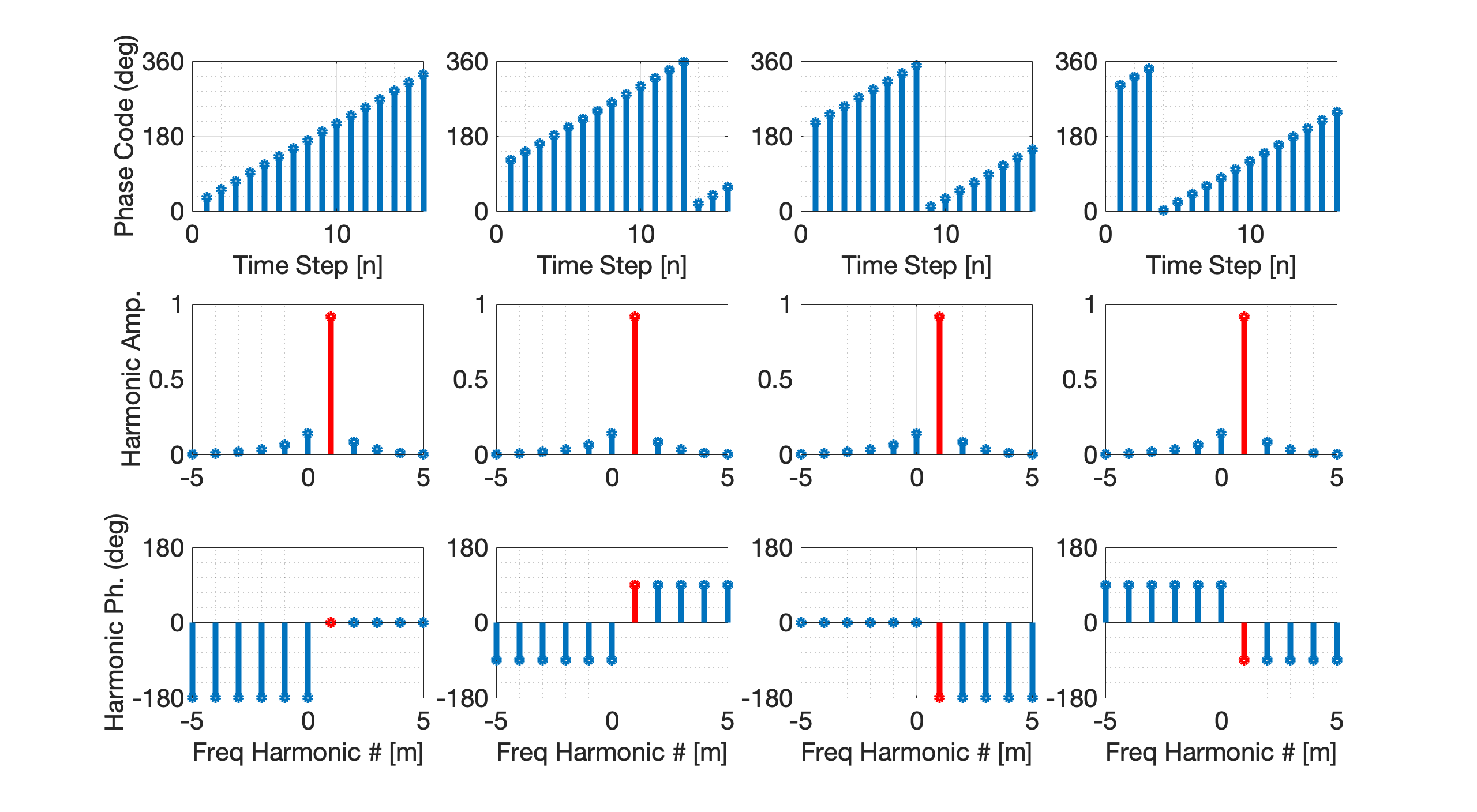}
  \caption{The phase of the GFH can be tuned tuned to controlled for communication modulation by time-shifting the meta-atom phase coding sequence using a circular shift. The example shown in this figure demonstrates the $m = +1$ GFH phase shifted to $-180$, $-90$, $0$, and $+90$ deg by controlling the time-shift of the phase coding sequence.}
  \label{fig:harmonic2}
\end{figure}

To implement FD-IM techniques, such as SIM, we need the ability to generate frequency harmonic beams using our R-MTS transceiver. In addition to the EM simulation results, we use \eqref{eqn:f100}-\eqref{eqn:f101} to analyze the R-MTS's harmonic generation response to different surface reflection phase codings. The time-varying R-MTS serves as a non-linear spatial mixer to manipulate frequency harmonics as described in Section~\ref{sec:R-MTS-Harmonic-Generation}. Results of this analysis are shown in Figs.~\ref{fig:harmonic1}-\ref{fig:harmonic2}. By changing the slope of the meta-atom phase coding between each time step during the modulation signaling interval, we control the index $m$ of the reflected frequency harmonic. In the most simple form, FD-IM is performed by the R-MTS transceiver using SIM-OOK (described in Section~\ref{sec:SIM-OOK}) by using the harmonic index $m$ to signal the data bits. The amplitude of each reflected frequency harmonic is modulated by controlling the range of phases used in the phase coding. If the range of phases used in a signaling interval is $<360$ deg, the amplitude of the generated frequency harmonic (GFH) will be reduced.

To modulate additional $M$-ary signal constellation bits using FD-IM, we also need the capability to independently modulate the phase of each GFH. We achieve phase tuning of GFHs by time-shifting the meta-atom phase coding sequence using a circular shift. Our numerical simulations showing the phase tuning of GFHs is shown in Figure~\ref{fig:harmonic2}. In this example, we show that by circular shifting the phase coding sequence by $L/4$, the phase of the GFH shifts by 90 degrees. Figure~\ref{fig:harmonic2} demonstrates the $m = +1$ GFH phase shifted to $-180$, $-90$, $0$, and $+90$ deg by controlling the time-shift of the phase coding sequence. From this numerical experiment data, we derive the relationship between the relative GFM phase and the length of the circular time shift in terms of number of time-steps $n$ as 
\begin{equation}
    n_{\textrm{shift}} = \frac{\phi_{\textrm{des}} L}{360 \: \mathrm{deg}}, 
\end{equation}
where $\phi_{\textrm{des}}$ is the desired phase shift of the GFH in degrees. Using this MTS-based time-varying phase coding technique, we are able to direct modulate the carrier wave to programmable generate reflected frequency harmonics of index $m$, with specified harmonic spacing, and a tailored phase of the frequency harmonic. These are the EM building blocks to perform SIM-OOK, SIM, OFDM-IM, and MIMO-OFDM-IM directly using the R-MTS transceiver as described in Section~\ref{sec:IM-techniques}. A conceptual illustration of the R-MTS transceiver implementation of OFDM-IM is shown in Figure~\ref{fig:MTS_ConceptualGraphic_ofdmim_revA}.

\begin{figure}[t]
  \centering
  \includegraphics[width=1.0\columnwidth]{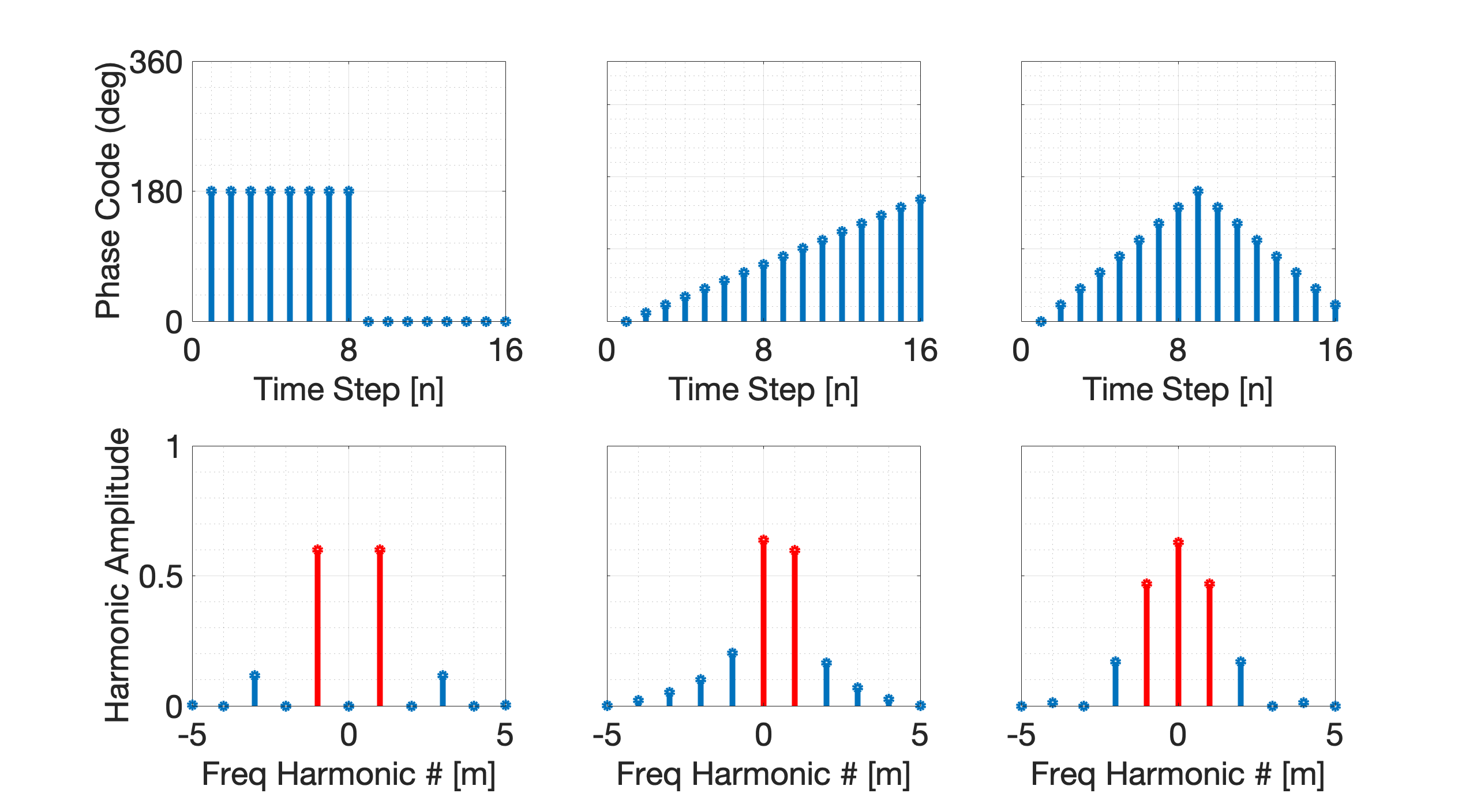}
  \caption{Examples of multiple simultaneous frequency harmonic. R-MTS harmonic generation from time-varying phase modulation.}
  \label{fig:harmonic3}
\end{figure}

The R-MTS transceiver can also generate multiple simultaneous harmonics using a proper phase coding. Three examples of multi-harmonic generation phase codings are shown in Figure~\ref{fig:harmonic3}. By temporally modulating the meta-atom phase in the correct programmable sequence, power from the carrier wave at $f_{c}$ can be translated to and allocated between the frequencies of interest for SIM-based wireless communication. In the spatial domain, GSM extends conventional SM by transmitting from multiple spatial indices (Tx antennas or RIS sub-apertures) simultaneous to increase data throughput as described in Section~\ref{sec:IM-techniques}. We extend this concept to the frequency domain by using the R-MTS time-varying phase coding to generate multiple simultaneous frequency subcarriers. In this manner, the R-MTS transceiver is able to increase data throughput while performing OFDM-IM by simultaneously transmitting bits using multiple subcarriers, where the index of the subcarriers and each subcarrier's $M$-ary constellation symbol are used to convey data for wireless communication.

\subsection{BER performance}
\begin{figure*}
\begin{center}
\noindent
  \includegraphics[width=1.0\textwidth]{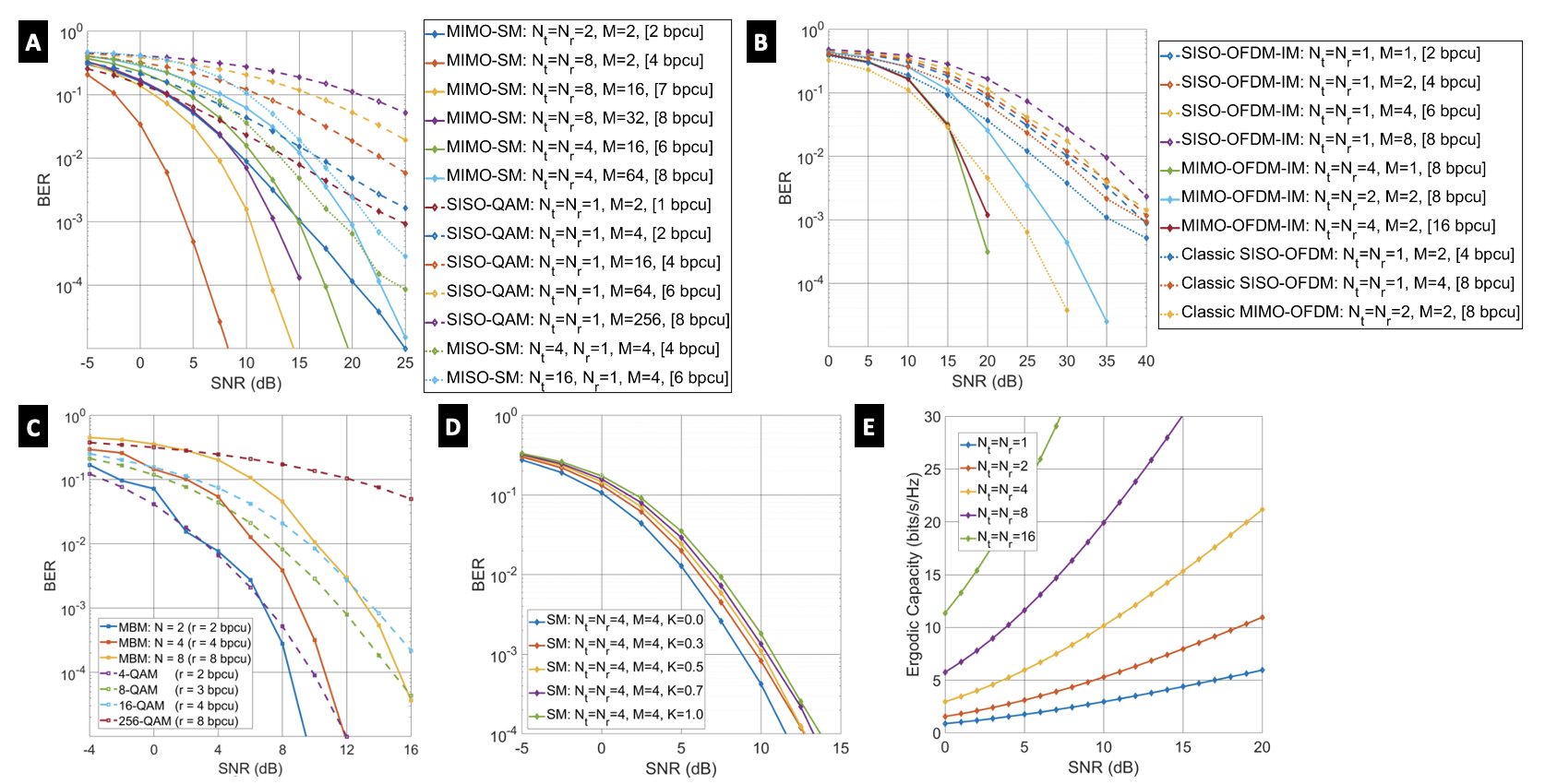}
  \caption{(a) Performance comparison of simulated bit error rate (BER) vs. signal-to-noise ratio (SNR) performance for SM and classical QAM systems under Rayleigh fading channels using an MLD. System variations are compared in terms of bits per channel use (bpcu). (b) Performance comparison of simulated BER vs. SNR performance for MIMO/MISO/SISO-OFDM-IM systems and classical MIMO/SISO-OFDM systems under Rayleigh fading channels using an MLD. (c) Simulated bit error rate (BER) performance for SIMO-MBM and classical SIMO $M$-QAM systems with $N_{r}=4$ receive antennas under uncorrelated AWGN channels. $N$ is the number of bits transmitted at each signaling interval. (d) BER performance of an SM system under different Rician fading conditions. We vary the Rician $k$ factor from 0 to 1 to simulate different fading channel conditions. The Rayleigh fading channel is equivalent to $K=0$. (e) Simulated ergodic channel capacity for $N_{t}=N_{r}= 1$ to $16$ systems under a Rayleigh fading channel.  
  }\label{fig:Fig1LabelBER}
\end{center}
\end{figure*}

To demonstrate the benefit of MTS-based IM systems, we performed bit error rate (BER) simulations and compared the IM-based results to traditional non-IM systems. The advantages for MTS-based IM systems is shown in Figures~\ref{fig:Fig1LabelBER}a-c. In both the spatial and frequency domains, the performance benefit of the IM system increases as $N_{t}$ and $N_{r}$ increase. 
Figure~\ref{fig:Fig1LabelBER}d shows the impact of the Rician or Rayleigh channel fading conditions for SM. Figure~\ref{fig:Fig1LabelBER}e shows simulated ergodic channel capacity for $N_{t}=N_{r}= 1$ to $16$ systems under a Rayleigh fading channel. As expected, channel capacity increases as the number of antennas in the MIMO system increases. By increasing the electrical size of the R-MTS (RIS/LIS), greater channel capacities can be supported in future networks.


\section{Summary}
\label{sec:summ}
We demonstrated the concept of using R-MTSs to simplify communication system architectures and support next-generation IM waveforms including SM and SIM/OFDM-IM. The MTS-based integrated antenna-transceiver architecture provides compelling advantages of low RF complexity, cost, and power consumption compared to current phased array antennas and heterodyne transceiver architectures. Additionally, the complexity and cost advantages of R-MTS antennas over traditional phased arrays scale with increasing aperture size. This makes direct modulation R-MTS arrays an attractive solution for IM, massive MIMO, and shared-aperture multi-beam antennas in 5G and 6G wireless networks.

\bibliographystyle{IEEEtran}
\bibliography{main}

\end{document}